\documentclass{jfm}

\usepackage{natbib}
\usepackage{amsmath}
\usepackage{amsfonts}
\usepackage{amssymb}
\usepackage[utf8]{inputenc}
\usepackage[T1]{fontenc}

\newcommand{\imagunit}{\mathrm{i}}
\newcommand{\diffunit}{\mathrm{d}}

\title{Coherent structure detection and the inverse cascade mechanism in two-dimensional Navier-Stokes turbulence}
\author{Jiahan Wang\aff{1}, Jörn Sesterhenn\aff{2} 
        \and Wolf-Christian Müller\aff{1}\corresp{wolf-christian.mueller@tu-berlin.de}
}
\affiliation{\aff{1}Zentrum f\"ur Astronomie und Astrophysik, ER 3-2, Technische Universit\"at Berlin, Hardenbergstr. 36a, 10623 Berlin, Germany
\aff{2}Technische Mechanik und Strömungsmechanik, Universit\"at Bayreuth, Universitätsstrasse 30, 95440 Bayreuth, Germany
}

\begin{document}
\maketitle
\begin{abstract}
Coherent structures in two-dimensional Navier-Stokes turbulence are
ubiquitously observed in nature, experiments and numerical
simulations. The present study conducts a comparison between several
structure detection schemes based on the Okubo-Weiss criterion, the
vorticity magnitude, and Lagrangian coherent structures (LCSs),
focusing on the inverse cascade in two-dimensional hydrodynamic
turbulence. A recently introduced vortex scaling phenomenology
[B. H. Burgess, R. K. Scott, J. Fluid Mech., 811:742--756, 2017]
allows the quantification of the respective thresholds required by these
methods based on physical properties of the flow. The resulting improved
comparability allows to identify characteristic relative
differences in the detection sensitivity between
the employed structure detection techniques. 
With respect to the inverse cascade of energy, 
coherent structures contribute, as expected, substantially less to the cross-scale flux
than the residual incoherent parts of the
flow although the energetically dominant coherent structures 
lead to an important large-scale deformation of the energy spectrum.
This cascade inactivity can be understood by an increased misalignment of
strain-rate and subgrid stress tensors within coherent structures. At
the same time, the structures exhibit strong and localised nonlinear
cross-scale interactions that appear to stabilize them.  We quantify
and interpret the resulting shape preservation of coherent structures
in terms of a 
multi-scale gradient approach [G. L. Eyink, J. Fluid Mech.,
549:191--214, 2006] as the depletion of strain rotation 
and vorticity gradient stretching while the dynamics of the residual fluctuations are
consistent with the vortex thinning picture.
\end{abstract}

\section{Introduction}\label{section1}
The formation of structures in
turbulent flows is ubiquitously observed in nature for example in
atmospheric flows or oceans.  In two-dimensional hydrodynamic turbulence,
the structure formation process is associated with the  inverse cascade of
kinetic energy,
tranferring this quantity from smaller to ever larger spatial scales. This
well-known phenomenon has already been predicted in the seminal papers of
\citet{Kraichnan1967a}, \citet{Leith1968} and \citet{Batchelor1969} (KLB) 
and observed in numerous simulations
\citep[e.g.][]{Lilly1971,FrischSulem1984,MaltrudVallis1993,BoffettaMusacchio2010}
and experiments
\citep[e.g.][]{Rutgers1998,ParetTabeling1998,ChenEckeEyinkRiveraWanXiao2006}.
We have chosen this physical system for the present work since it allows to study naturally emergent
coherence typically appearing in the form of structurally rather simple vortices or combinations of those.
These vortical structures are embedded in a statistically isotropic and turbulent two-dimensional flow which is conveniently
accessible to direct numerical simulation (DNS) and measurement.

An intuitive and commonly accepted defining characteristic of 
coherence is persistence 
for a finite time, which leaves  room for more detailed specification.
A mathematically unique
definition would not only be beneficial for fluid mechanics research but
also for related disciplines such as astrophysics.
There, the problem of
the non-universality regarding vortex identification has been
pointed out by \citet{CaniveteCuissaSteiner2020} and \citet{YadavCameronSolanki2021} 
with respect to studies of the solar atmosphere.

A number of methods for the detection of coherent structures exist 
that are often built on different specifications of coherence. 
In fact, most of the comparative studies in the literature focus on a specific class
of coherence specification, e.g. \citet{JeongHussain1995} studied various
vortex criteria, \citet{HadjighasemFarazmandBlazevskiFroylandHaller2017} 
compared different techniques for Lagrangian coherent structure (LCS)
identification and \citet{TairaBruntonDawsonRowleyColoniusMcKeonSchmidtGordeyevTheofilisUkeiley2017}
discussed numerous mode decomposition methods.
A comparison and meaningful evaluation of different detection strategies and of their respective
coherence specification requires fiducial physical properties of the considered flow which can be related to the detected
structures.

In the present work, we consider three coherence
detection schemes -- two vorticity based, one of Lagrangian type -- that use the Okubo-Weiss criterion
\citep{Okubo1970,Weiss1991}, the vorticity magnitude, and the
finite-time Lyapunov exponent (FTLE) field determining LCSs
\citep{HallerYuan2000,Haller2015}.
We compare the detection results by investigating the physical properties of
the coherent and the residual (non-coherent) structures in the two-dimensional turbulent system.

The work of \citet{Ouellette2012} follows a similar approach,
which has led to several experimental studies
\citep[see][]{LiaoOuellette2013,KelleyAllshouseOuellette2013}. 
The low Reynolds numbers attained in these two-dimensional experiments ($Re = 185$ \citep{LiaoOuellette2013} 
and $Re = 220$ \citep{KelleyAllshouseOuellette2013}) however do not allow for, e.g., an
adequate investigation of cross-scale
turbulent interactions in the framework of the KLB similarity ansatz.
This motivates the use of DNSs in the present
investigation.

More specifically, we consider spectral nonlinear cross-scale fluxes of energy,
as well as their scale-filtered correspondents in configuration space.
We also compare with previous results based on a multi-scale gradient ansatz
and employ
a recently proposed vortex scaling phenomenology for two-dimensional Navier-Stokes turbulence 
\citep{BurgessScott2017,BurgessScott2018} which applies dimensional arguments 
to impose physically motivated constraints onto coherent vortices. 
The results obtained by these theoretical approaches serve as  physical reference
points for the comparison of structure detection techniques and the interpretation of their
results in relation to the inverse turbulent cascade of energy.

This paper is structured as follows: 
Section \ref{section2} presents the decomposition of the
flow into coherent and residual contributions. Section \ref{section3} 
briefly introduces coherent structure specifications. Section \ref{section4}
describes the applied diagnostics and theoretical concepts. Section \ref{section5} presents
numerical methods and the parameters used for simulations and
analysis. The main results are presented in section
\ref{section6}. A conclusion is given in section \ref{section7}.

\section{Physical model and flow decomposition}\label{section2}
We consider Navier-Stokes turbulence 
on a two-dimensional $2\pi$-periodic square of size $A$ governed by the differential equations,
\begin{align}
 \partial_t \boldsymbol{u} + (\boldsymbol{u} \bcdot \nabla) \boldsymbol{u} &= -\nabla p + \nu \nabla^2 \boldsymbol{u}, \label{eq:navier_stokes1} \\
 \nabla \bcdot \boldsymbol{u} &= 0, \label{eq:navier_stokes2}
\end{align}
where $\boldsymbol{u} = (u_x, u_y)$, $p$ and $\nu$ are the velocity, pressure
and kinematic viscosity, respectively.
The kinetic energy per unit mass $E = (1/2A) \int_A \boldsymbol{u}^2 dA$ and
the enstrophy $\Omega = (1/2A) \int_A \omega^2 dA$ defined with the vorticity $\omega = \partial_x u_y -
\partial_y u_x$, are inviscid invariants 
in a two-dimensional configuration. The kinetic energy exhibits
an inverse cascade, transferring energy from small to large length
scales, contrary to the enstrophy, which exhibits a direct cascade. 

We employ a decomposition of the total vorticity field into a coherent part,
$\omega_c$, and a residual/incoherent contribution, $\omega_r$,
\citep[cf.][]{Ohkitani1991} to carry out the analysis of different
schemes for coherence detection (cf. section \ref{section3} below):
\begin{align}
\omega &= \omega_c + \omega_r, \nonumber \\
\omega_c(\boldsymbol{x}) &= \begin{cases}
      \omega(\boldsymbol{x}), &\epsilon(\boldsymbol{x}) \geq \epsilon_{thr}, \\
      0, &\epsilon(\boldsymbol{x}) < \epsilon_{thr},
\end{cases} \label{eq:decomposition_coherent} \\
\omega_r(\boldsymbol{x}) &= \begin{cases}
      0,  &\epsilon(\boldsymbol{x}) \geq \epsilon_{thr}, \\
      \omega(\boldsymbol{x}), &\epsilon(\boldsymbol{x}) < \epsilon_{thr}.
\end{cases} \label{eq:decomposition_residual}
\end{align}
Based on the particular specification of coherence, a physical
characteristic of the flow, $\epsilon(\boldsymbol{x})$, serves as an indicator
of this property, turning the detection into a thresholding procedure
with a fixed threshold, $\epsilon_{thr}$. 
In order to improve comparability of different detection schemes, it 
is important to gauge their thresholds with respect to a
physical property of the flow (see section \ref{section3_4} below).
Technical details of the decomposition are pointed out
in appendix \ref{appendix1_1}. The coherent
velocity field, $\boldsymbol{u}_c = (u_{c,x}, u_{c,y})$, and the residual velocity
$\boldsymbol{u}_r = (u_{r,x}, u_{r,y})$, are approximated by inverting $\boldsymbol
\omega_{c/r}=\nabla\times\boldsymbol{u}_{c/r}$ in Fourier space, a procedure
symbolically represented by the operator $\nabla\times\nabla^{-2}$,
which similarly has been employed in several related works
\citep[see][]{BenziPaladinPatarnelloSantangeloVulpiani1986,BenziPatarnelloSantangelo1988,Borue1994,
OkamotoYoshimatsuSchneiderFargeKaneda2007,YoshimatsuKondoSchneiderOkamotoHagiwaraFarge2009,
Vallgren2011,BurgessScott2018}:
\begin{align}
\boldsymbol{u}_{c/r} &= -\nabla \times (\nabla^{-2} \boldsymbol{\omega}_{c/r}), \nonumber \\
\boldsymbol{u} &= \boldsymbol{u}_{c} + \boldsymbol{u}_{r}\,. \label{eq:biot_savarts_law}
\end{align}
This enables straightforward access to various decomposed turbulent
fields and related quantities such as the Fourier spectrum of kinetic
energy per unit mass. It is defined as $E(k) = \sum\nolimits_k \left|
\hat{\boldsymbol{u}}(\boldsymbol{k}) \right|^2 / 2$ with the wavevector
$\boldsymbol{k}=(k_x,k_y)$ and for the respective length scale $\ell \sim
k^{-1}$. Fourier-transformed quantities are denoted by
\lq$\,\string^\,$\rq\ and the sum over all wavevectors located
on a wavenumber shell, $k \leq \left| \boldsymbol{k} \right| < k+1$,  is
indicated as $\sum\nolimits_k$. According to the KLB phenomenology, the
spectrum possesses scaling properties for the inverse kinetic
energy and direct enstrophy cascade ranges, which are
\begin{align}
 E(k) &\sim \epsilon_I^{2/3} k^{-5/3}, \quad k \ll k_f, \\
 E(k) &\sim \eta_I^{2/3} k^{-3}, \quad k \gg k_f,
\end{align}
with $k_f$ the forcing wavenumber at which energy and enstrophy are
injected with an injection rate $\epsilon_I$ or $\eta_I = k_f^2
\epsilon_I$, respectively.
Here, we are interested in a decomposition of the kinetic energy
spectrum as
\begin{align}
 E(k) = E_{c}(k) + E_{r}(k) + E_{cr}(k), \label{eq:decomposed_energy_spectrum}
\end{align}
where $E_{c}(k) = \sum\nolimits_k \left| \hat{\boldsymbol{u}}_c(\boldsymbol{k})
\right|^2 / 2$ and $E_{r}(k) = \sum\nolimits_k \left|
\hat{\boldsymbol{u}}_r(\boldsymbol{k}) \right|^2 / 2$ are associated to spectral
contributions from purely coherent and residual regions, respectively,
and $E_{cr}(k) = \sum\nolimits_k \Real[\hat{\boldsymbol{u}}^{*}_c(\boldsymbol{k}) \bcdot
\hat{\boldsymbol{u}}_r(\boldsymbol{k})]$ the spectrum resulting from
mixed-contributions of coherent and residual parts, with \lq$\,*\,$\rq\ denoting
the complex conjugate.

In the following, we introduce the identification schemes considered
for $\epsilon(\boldsymbol{x})$ in equations \eqref{eq:decomposition_coherent}
and \eqref{eq:decomposition_residual} and the choice of the corresponding threshold
values $\epsilon_{thr}$.

\section{Coherence specifications}\label{section3}
In general, schemes for the detection of coherent structures can be grouped into
several categories including threshold methods, modal decomposition methods
such as proper orthogonal decomposition (POD) 
\citep{HolmesLumleyBerkoozRowley_TurbulenceCoherentStructuresDynamicalSystemsAndSymmetry_2012},
dynamic mode decomposition (DMD) \citep{RowleyMezicBagheriSchlatterHenningson2009,Schmid2010}
or spectral proper orthonal decomposition (SPOD)
\citep{TowneSchmidtColonius2018},
and wavelet methods
\citep[see][]{OkamotoYoshimatsuSchneiderFargeKaneda2007,
YoshimatsuKondoSchneiderOkamotoHagiwaraFarge2009,FargeSchneider2015}. In
this work, we focus on threshold methods, in particular the
approaches based on the Okubo-Weiss (OW) criterion \citep{Okubo1970,Weiss1991}, 
the vorticity magnitude (VM), and the LCSs \citep{HallerYuan2000,Haller2015}.
These schemes are straightforwardly employed using equations
\eqref{eq:decomposition_coherent} and
\eqref{eq:decomposition_residual}.
The thresholds for the VM and for the LCS based structure detection 
are chosen with the help of vortex scaling 
(see section \ref{section3_4} below).

\subsection{Okubo-Weiss (OW)/$Q$-criterion}\label{section3_1}
A frequently applied quantity for structure identification is the
Eulerian velocity gradient tensor $\nabla \boldsymbol{u}$, which is often
investigated in decomposed form $\nabla \boldsymbol{u} = \mathsfbi{S} + \mathsfbi{W}$,
with $\mathsfbi{S} = (\nabla \boldsymbol{u} + (\nabla \boldsymbol{u})^T)/2$ the symmetric
strain-rate tensor and $\mathsfbi{W} = (\nabla \boldsymbol{u} - (\nabla
\boldsymbol{u})^T)/2$ the skew-symmetric spin tensor. 
The usage of invariants of $\nabla \boldsymbol{u}$, 
e.g. the eigenvalues or the trace, and of its tensor decomposition
have led to numerous identification schemes, see, e.g. \citep{HuntWrayMoin1988,ChongPerryCantwell1990,JeongHussain1995,
HuaKlein1998,ZhouAdrianBalachandarKendall1999,ChakrabortyBalachandarAdrian2005}.
However, all of these methods face the problem of 
\textit{objectivity} \citep{Haller2005,HallerHadjighasemFarazmandHuhn2016}, 
i.e. they lack
invariance under certain transformations of the frame of reference
which combine rotation and translation. Thus for the sake of
simplicity, we restrict ourselves to 
the well-known $Q$-criterion \citep{HuntWrayMoin1988}, whose two-dimensional
equivalent resembles the OW criterion. It is defined as
\begin{align}
 Q = \frac{1}{2} \left( \left| \mathsfbi{W} \right|^2 - \left| \mathsfbi{S} \right|^2 \right) = \frac{1}{2} \left( \frac{1}{2} \omega^2 - \left| \mathsfbi{S} \right|^2 \right),
\end{align}
where for $Q > 0$ vortex dominated/elliptical regions and for $Q < 0$
strain dominated/hyperbolic regions are detected. Thus,
$\epsilon(\boldsymbol{x}) = Q(\boldsymbol{x})$ and $\epsilon_{thr}=0$ are set in
equations \eqref{eq:decomposition_coherent} and
\eqref{eq:decomposition_residual}.

\subsection{Vorticity magnitude (VM)}\label{section3_2}
Coherent structures in two-dimensional flows are often most clearly
visible in the spatial distribution of the vorticity. Thus, an
intuitive approach is to set $\epsilon(\boldsymbol{x}) = \left| \omega(\boldsymbol{x})
\right|$ in equations \eqref{eq:decomposition_coherent} and
\eqref{eq:decomposition_residual}. Furthermore, the vorticity is
closely connected to the Lagrangian-averaged vorticity deviation
(LAVD) method \citep{HallerHadjighasemFarazmandHuhn2016}, which is an objective
detection criterion.

\subsection{Lagrangian coherent structures (LCSs)}\label{section3_3}
LCSs take the evolution of the flow field into account by
determining the pair-dispersion characteristic of passively advected
Lagrangian tracers \citep{HallerYuan2000,Haller2015}. Thus, they reveal
structures in the flow, which are neither captured by the vorticity
$\omega$ nor variants of the velocity gradient tensor $\nabla
\boldsymbol{u}$. To this end, the flowmap $\boldsymbol{F}_{t_0}^{t}(\boldsymbol{x}_0) =
\boldsymbol{x}(t;t_0,\boldsymbol{x}_0)$ is considered, with $\boldsymbol{x}_0 = (x_0,y_0)$ 
the initial position at time $t_0$. The detection of LCSs can be realised 
by determining the finite-time Lyapunov exponent (FTLE) field which is given by
\begin{align}
\Lambda_{t_0}^{t}(\boldsymbol{x}_0) = \frac{1}{T_{eddy}} \log \sqrt{\lambda^C_2(\boldsymbol{x}_0)}, \label{eq:FTLE}
\end{align}
with $\lambda^C_{2}$ the largest eigenvalue of the Cauchy-Green strain
tensor $\boldsymbol{C}_{t_0}^t(\boldsymbol{x}_0) = \left[ \nabla
\boldsymbol{F}_{t_0}^t(\boldsymbol{x}_0) \right]^T \nabla
\boldsymbol{F}_{t_0}^t(\boldsymbol{x}_0)$. The FTLE is interpreted as a local measure
of stretching and can be calculated forward and backward in time.
Thus, the values are set to $\epsilon(\boldsymbol{x}) =
\Lambda_{t_0}^{t_0+T_{eddy}}(\boldsymbol{x})$ for the forward-in-time and
$\epsilon(\boldsymbol{x}) = \Lambda_{t_0}^{t_0-T_{eddy}}(\boldsymbol{x})$ for the
backward-in-time case. Please note, that for the FTLE case the roles of
$\omega_c$ and $\omega_r$ are switched in equations
\eqref{eq:decomposition_coherent} and
\eqref{eq:decomposition_residual}, meaning that small FTLE values
correspond to coherent regions, contrary to the VM
$\left| \omega(\boldsymbol{x}) \right|$. This is because large FTLE values
isolate coherent regions as illustrated in figures
\ref{fig:various_physical_quantities_run2_4} (c) and (d).
To our knowledge no condition exists for the flowmap integration
time. Hence, we suggest setting it to the large-eddy turnover time
$T_{eddy}$ according to section \ref{section5}, which is typically the
longest characteristic correlation time scale of the system. Further
numerical details for the FTLE calculation are discussed in appendix
\ref{appendix1_2}.

Although more refined LCS approaches exist, for our purposes the FTLE yields
sufficient insight into the flow physics, 
as high-valued FTLE regions, which are visually perceived as sharp ridges in the flow,
are supposed to materially separate dynamically distinct domains 
with different transport characteristics. For example, these domains mark 
areas of zero cross-scale energy fluxes in low Reynolds number systems
\citep[cf.][]{KelleyAllshouseOuellette2013}. Furthermore, forward-in-time FTLE 
(f-FTLE) ridges are associated with repelling LCSs and 
backward-in-time FTLE (b-FTLE) ridges to attracting LCSs, 
indicating stable and unstable manifolds in
the flow in the sense of dynamical systems theory.

\subsection{Determining the threshold: vortex scaling}\label{section3_4}
Two of the three detection schemes considered here include free threshold parameters which
complicate a meaningful comparison of the detection methods and the physical interpretation of the detection results. 
In order to achieve comparability between the three coherence specifications,
the VM and LCS schemes are gauged by making use of the above-mentioned vortex scaling phenomenology.

This model, which we briefly summarize here for completeness, provides a physically motivated
diagnostic signature which  we use as a reference for the highly non-trivial threshold choice of $\epsilon_{thr}$
in equations \eqref{eq:decomposition_coherent} and
\eqref{eq:decomposition_residual}. The phenomenology characterizes coherent
structures by their vortex area $A$ in configuration space
instead of the classical wavenumber dependence in Fourier
space. Therefore, a time-dependent vortex number density distribution
$n(A,t)$ is defined, which yields the number of coherent vortices per
unit area for a certain vortex area $A$ at time $t$. The model is
based on the first three moments of $n\overline{\omega_v^2}$ with the
vortex intensity $\overline{\omega_v^2}$. They are {the vortex
energy $E_v$, vortex enstrophy $Z_v$ and vortex number $N_v$},
respectively. All three quantities are assumed to be approximately conserved 
during the spatial growth of an \lq average\rq\ vortex of area $A$. The number 
density is anticipated to follow a power-law $t^{\alpha_i} A^{-r_i}$ with 
exponents $\alpha_i$ and $r_i$ determined via the conservation of 
$E_v$, $Z_v$ and $N_v$. The range of areas is divided into a \textit{thermal
bath} regime $A_f \leq A < A_{-}$, an \textit{intermediate} scaling
regime $A_{-} < A < A_{+}$ and a \textit{front} of the vortex
population $A_{+} < A \leq A_{max}$, respectively, where $A_{-}$ and
$A_{+}$ are transitional areas, $A_f$ the forcing-scale area, and
$A_{max}$ the maximum vortex area.

In this model, the thermal bath is associated with the equilibration 
of the flow with the continuous forcing, which injects energy at a 
constant rate generating small-scale vorticity. This leads to an
$A$-independent flux of $E_v$ in $A$-space. The intermediate scaling 
regime consists of a self-similar distribution of vortex sizes. It is 
assumed that the enstrophy lost through filament shedding during merger and 
aggregation processes is replaced by the enstrophy injection such that the vortex enstrophy
$Z_v$ is also approximately conserved. In the front regime, vortices are expected to be
large and distant from each other, such that merging events rarely
occur. Thus, approximately conserving the vortex number $N_v$. Based on these
conservation assumptions, the scaling laws of the number density for
varying area regimes are derived as
\citep[see][]{BurgessScott2017,BurgessScott2018}
\begin{align}
n(A,t) \sim \begin{cases}
      A^{-3}, & A_f \leq A < A_{-}, \\
      t^{-1} A^{-1}, & A_{-} < A < A_{+}, \\
      t^{5} A^{-6}, & A_{+} < A \leq A_{max}. \label{eq:three_part_number_density}
    \end{cases}
\end{align}

We take the best achievable agreement with the three regime subdivision
\eqref{eq:three_part_number_density} as a reference to gauge
the threshold values in \eqref{eq:decomposition_coherent} and
\eqref{eq:decomposition_residual}. Please note that this qualitative level
of agreement mainly relies on the assumption that the emergence and the evolution of coherence
are asymptotically self-similar for sufficiently large scale-separation between the regions of the forcing and the large scales of the
system under consideration. This can only be fulfilled up to a rather modest approximate level in turbulence DNS. 
In the present work, the scaling exponents are considered relative to each other. Thus,
their absolute numerical values 
are not of principal importance to the investigation. They are nevertheless mentioned above for
completeness.

\section{Diagnostic methods for the inverse cascade}\label{section4}
The inverse cascade of kinetic energy corresponds to a
cross-scale energy flux of which we distinguish
coherent and residual contributions 
from three perspectives:
$(i)$ spectrally in Fourier space, $(ii)$ scale-filtered in
configuration space which combines the aspects of spatial scale and position, 
and $(iii)$ via a multi-scale gradient (MSG) approach \citep{Eyink2006b} which adds 
scale locality and the differentiation between involved physical processes.

\subsection{Spectral flux}\label{section4_1}
The temporal evolution of the energy
spectrum is straightforwardly obtained from the Navier-Stokes equations 
\eqref{eq:navier_stokes1} and \eqref{eq:navier_stokes2} as:
\begin{align}
 \partial_t E(k) + T(k) = D(k) + F(k),
\end{align}
with the nonlinear transfer term
$T(k) = \sum\nolimits_k \Real [\hat{\boldsymbol{u}}^{*}(\boldsymbol{k}) 
\bcdot \widehat{(\boldsymbol{u} \bcdot \nabla \boldsymbol{u})}(\boldsymbol{k})]$. 
Kinetic energy is provided to the flow by a forcing term $+\boldsymbol{f}_u$ on 
the right-hand side of the Navier-Stokes momentum balance \eqref{eq:navier_stokes1}. 
Thus, the energy source term is determined as $F(k) = \sum\nolimits_k \Real[
\hat{\boldsymbol{u}}^{*}(\boldsymbol{k}) \bcdot \hat{\boldsymbol{f}}_u(\boldsymbol{k})]$, 
which is equivalent to the energy injection rate $\epsilon_I$ when summed
over all Fourier wavenumbers. In order to allow for a statistically stationary 
state, kinetic energy accumulating at the largest length scales of the flow
due to the inverse cascade has to be continuously extracted from the system.
For this purpose a  large scale damping term $-d_{\alpha} \boldsymbol{u}$ is added 
to the right-hand side of equation \eqref{eq:navier_stokes1}.
The energy sink $D(k) = D_{\nu}(k) + D_{\alpha}(k)$
is split into two dissipative contributions, where $D_{\nu}(k) = -2 \nu k^2 E(k)$ 
is the viscous dissipation active on small length scales 
and $D_{\alpha}(k) = -2 d_{\alpha} E(k)$ introduces friction active on large length scales. 
These terms are equivalent to the energy dissipation rate on 
viscous scales $\epsilon_{\nu}$ and on large scales $\epsilon_{\alpha}$,
respectively, when summed over all wavenumbers.
Details on the numerical implementation of $+\boldsymbol{f}_u$ and $-d_{\alpha} \boldsymbol{u}$
are given in the text around equation \eqref{eq:navier_stokes_numerical_implementation} in section \ref{section5} .

The spectral cross-scale energy flux $Z(k)$ is obtained by summing 
the transfer term over consecutive shells
\begin{align}
 Z(k) = \sum_{k'=0}^k T(k') = \sum_{k'=0}^k \sum\nolimits_{k'} \Real\left[\hat{\boldsymbol{u}}^{*}(\boldsymbol{k}) \bcdot \widehat{(\boldsymbol{u} \bcdot \nabla \boldsymbol{u})}(\boldsymbol{k})\right], \label{eq:spectral_flux}
\end{align}
and corresponds to the flux of energy from scales smaller than $k$
to scales larger than $k$. 
The influence of the coherent and residual contributions with regard to the cascade
mechanism is measured by the decomposition
\begin{align}
 Z(k) &= \sum_{\alpha,\beta,\gamma \in \{c,r\}} Z_{\alpha,\beta,\gamma}(k), \\
 Z_{\alpha,\beta,\gamma}(k) = \sum_{k'=0}^k T_{\alpha,\beta,\gamma}(k') &= \sum_{k'=0}^k \sum\nolimits_{k'} \Real\left[\hat{\boldsymbol{u}}_{\alpha}^{*}(\boldsymbol{k}) \bcdot \widehat{(\boldsymbol{u}_{\beta} \bcdot \nabla \boldsymbol{u}_{\gamma})}(\boldsymbol{k})\right], \label{eq:decomposed_spectral_scale_flux}
\end{align}
which results in eight independent flux contributions. We investigate
the homogeneous fluxes originating from purely coherent
$Z_{c,c,c}(k)$ and residual components $Z_{r,r,r}(k)$, and the
mixed flux arising through coherent-residual interactions as
$Z_{cr}(k) = Z(k) - Z_{c,c,c}(k) - Z_{r,r,r}(k)$.

\subsection{Spatial flux distribution}\label{section4_2}
A complementary formulation of the cross-scale energy flux which captures its 
local structure in configuration space and which enables a detailed analysis 
regarding its spatial distribution is obtained by a
scale-filter approach, cf., e.g., \citep{Ouellette2012}.
For the $i$-th component of the velocity vector
the filter operation at a length scale $\ell$ is given by
\begin{align}
\overline{u}_i(\boldsymbol{x}) = \int G_{\ell}(\boldsymbol{r}) u_i(\boldsymbol{x}+\boldsymbol{r}) \diffunit^2 \boldsymbol{r},
\end{align}
where we choose $G_{\ell}$ as a smooth, non-negative, spatially well localised filter kernel
with unit integral. Because we are interested in the spatial distribution of the cross-scale flux, 
the locality aspect of the filter is crucial for the accurate localisation of 
flux contributions in configuration space. Hence, in this work we employ a
Gaussian filter: $\hat{G}_{\ell}(\boldsymbol{k})=\exp(-k^2 {\ell}^2/24)$ 
to achieve sufficient filter locality in Fourier space as well as in configuration space.
The temporal evolution of the filtered kinetic energy, 
$\overline{E} = \left| \overline{\boldsymbol{u}} \right|^2/2$, is given by \citep[see][]{Pope_TurbulentFlows_2000}
\begin{align}
 \partial_t \overline{E} + \partial_i \overline{q}_i = -\overline{\epsilon}_{\nu} - \overline{Z},
\end{align}
with $\partial_i$ the partial derivative of the $i$-th component,
where we use the Einstein summation convention. Additionally,
$\overline{q}_i = \overline{u}_i \overline{E} +
\overline{u}_j (\overline{p} \delta_{ij} + \overline{\tau}_{ij}
- 2 \nu \overline{S}_{ij})$ contains the 
nonlinear spatial transport and the viscous dissipation of the filtered
large-scale kinetic energy with $\delta_{ij}$ the Kronecker delta function,
$\overline{\epsilon}_{\nu} = 2 \nu \overline{S}_{ij}
\overline{S}_{ij}$ the viscous dissipation from the filtered velocity
field, and $\overline{Z} = - \overline{S}_{ij} \overline{\tau}_{ij}$ the spatial cross-scale
flux term representing the exchange of kinetic energy between the known filtered fields
and the fluctuations which have been depleted by the 
filter operation in the filtered numerical system. 
The flux term is an inner product of the 
(filtered) strain-rate tensor $\overline{S}_{ij} = 
(\partial_j \overline{u}_{i} + \partial_i \overline{u}_{j})/2$ 
and the subgrid stress tensor $\overline{\tau}_{ij} =
\overline{u_i u_j} - \overline{u}_i \overline{u}_j$, 
that expresses the stresses exerted by the depleted
fluctuations. Please note that
we are referring to the deviatoric (trace-free) stress term
$\mathring{\overline{\boldsymbol{\tau}}} = \overline{\boldsymbol{\tau}} - (1/2)
tr(\overline{\boldsymbol{\tau}}) \mathbb{I}$, with $tr$ the trace operator and
$\mathbb{I}$ the unit matrix. For the remainder we will write
$\overline{\boldsymbol{\tau}}$ instead of $\mathring{\overline{\boldsymbol{\tau}}}$.
The production term $\overline{Z}$ is equally understood as a spatial
cross-scale flux term. Note that choosing a sharp spectral filter instead of
a smooth Gaussian filter will lead to the equality between the spatial
average of the production term and the spectral flux in equation
\eqref{eq:spectral_flux} as $\langle \overline{Z}(\boldsymbol{x}) \rangle =
Z(k=2\pi/\ell)$, if the wavenumber $k$ is chosen according to the
filtering length scale $\ell$.

The strain-rate and stress tensor
are further decomposed into coherent, residual and mixed contributions
\begin{align}
\overline{Z}(\boldsymbol{x}) = \sum_{\alpha,\beta,\gamma \in \{ c, r \}} \overline{Z}_{\alpha,\beta,\gamma}(\boldsymbol{x}) = \sum_{\alpha,\beta,\gamma \in \{ c, r \}} \overline{\mathsfbi{S}}_{\alpha} : \overline{\boldsymbol{\tau}}_{\beta,\gamma},
\end{align}
with $\overline{\mathsfbi{S}}_{\alpha} = (\nabla \overline{\boldsymbol{u}}_{\alpha} + 
(\nabla \overline{\boldsymbol{u}}_{\alpha})^T)/2$ and $\overline{\boldsymbol{\tau}}_{\beta,\gamma} = 
\overline{\boldsymbol{u}_{\beta} \boldsymbol{u}_{\gamma}} - 
\overline{\boldsymbol{u}}_{\beta} \overline{\boldsymbol{u}}_{\gamma}$. 
We propose the following three-part decomposition
\begin{align}
 \overline{Z}(\boldsymbol{x}) = -\overline{\mathsfbi{S}} : \overline{\boldsymbol{\tau}} = & \underbrace{-\overline{\mathsfbi{S}}_c : \overline{\boldsymbol{\tau}}_{c,c}}_{\overline{Z}_c(\boldsymbol{x})} \underbrace{-\overline{\mathsfbi{S}}_r : \overline{\boldsymbol{\tau}}_{r,r}}_{\overline{Z}_r(\boldsymbol{x})} \nonumber \\
 & \underbrace{-\overline{\mathsfbi{S}}_c : \overline{\boldsymbol{\tau}}_{r,r} -\overline{\mathsfbi{S}}_r : \overline{\boldsymbol{\tau}}_{c,c} -\overline{\mathsfbi{S}}_c : (\overline{\boldsymbol{\tau}}_{c,r} + \overline{\boldsymbol{\tau}}_{r,c}) -\overline{\mathsfbi{S}}_r : (\overline{\boldsymbol{\tau}}_{c,r} + \overline{\boldsymbol{\tau}}_{r,c})}_{\overline{Z}_{cr}(\boldsymbol{x})}, \label{eq:decomposed_spatial_scale_flux}
\end{align}
where $\overline{Z}_c(\boldsymbol{x})$ consists of purely coherent and
$\overline{Z}_r(\boldsymbol{x})$ of purely residual contributions, and
$\overline{Z}_{cr}(\boldsymbol{x})$ is the flux contribution originating from
mixed interactions.

\begin{figure}
\centering
\captionsetup{width=\textwidth}
\includegraphics[width=\textwidth]{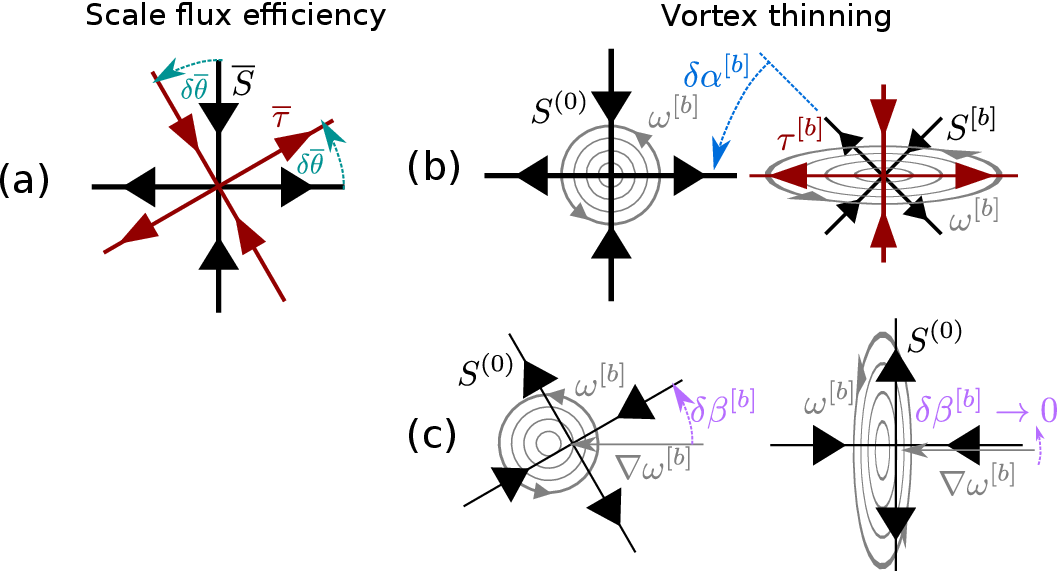}
\caption{Overview of different angles used in this work. (a): Angle $\delta \overline{\theta}$ 
between the strain-rate $\overline{\mathsfbi{S}}$ and subgrid stress tensor 
$\overline{\boldsymbol{\tau}}$. (b): Angle $\delta \alpha^{[b]}$ between the large-scale 
strain-rate tensor $\mathsfbi{S}^{(0)}$ and the band-pass filtered strain-rate tensor 
$\mathsfbi{S}^{[b]}$. Remade from \citep{Eyink2006b}. (c): Angle $\delta \beta^{[b]}$ 
between the contractile direction of the large-scale strain-rate tensor $\mathsfbi{S}^{(0)}$ 
and the band-pass filtered vorticity gradient vector $\nabla \omega^{[b]}$.}
\label{fig:rotation_angle_overview}
\end{figure}

Because the spatial cross-scale flux consists of an inner product of two
tensors, the analysis of angle alignments between tensor eigenframes
is possible. Thus, a polar decomposition leads to the following
expression for the total and decomposed fluxes
\citep[see][]{Eyink2006b,FangOuellette2016}
\begin{align}
 \overline{Z}(\boldsymbol{x}) &= -2 \overline{\sigma}(\boldsymbol{x}) \overline{\lambda}(\boldsymbol{x}) \cos(2 \delta \overline{\theta}(\boldsymbol{x})), \label{eq:polar_spatial_scale_flux} \\
 \overline{Z}_{c/r}(\boldsymbol{x}) &= -2 \overline{\sigma}_{c/r}(\boldsymbol{x}) \overline{\lambda}_{c/r}(\boldsymbol{x}) \cos(2 \delta \overline{\theta}_{c/r}(\boldsymbol{x})), \label{eq:decomposed_polar_spatial_scale_flux}
\end{align}
respectively. The positive eigenvalues of the strain-rate and subgrid
stress tensors are $\overline{\sigma}$ and $\overline{\lambda}$,
respectively, and the angle between their corresponding eigenvectors is
$\delta \overline{\theta}$ as illustrated in figure
\ref{fig:rotation_angle_overview} (a). The same definitions are used
for the eigenvalues and angles of coherent and residual parts, which
are indicated by the indices $c$ and $r$, respectively. The cosine of
the rotation angle between strain-rate and stress tensors, $\cos(2
\delta \overline{\theta})$, can be understood as an \textit{efficiency} of the
cross-scale energy transfer \citep{FangOuellette2016}. Therefore, a
detailed analysis of angle distributions from coherent $\delta
\overline{\theta}_{c}$ and residual parts $\delta
\overline{\theta}_{r}$ is conducted in section \ref{section6_1}.

The mixed cross-scale flux $\overline{Z}_{cr}$ 
in equation \eqref{eq:decomposed_spatial_scale_flux} is a very complex object due to the 
heterogeneous subgrid stress tensors,
$\overline{\boldsymbol{\tau}}_{c,r}$ and $\overline{\boldsymbol{\tau}}_{r,c}$, which are
not symmetric and thus not straightforward to interpret.
Only the sum of $\overline{\boldsymbol{\tau}}_{c,r} + \overline{\boldsymbol{\tau}}_{r,c}$
yields a symmetric stress quantity.
Thus, the mixed cross-scale flux contribution consists of a
sum of four different physical contributions: $(i)$ Exertion of residual
stress on coherent strain-rate, $(ii)$ exertion of coherent stress on
residual strain-rate, $(iii)$ exertion of mixed stress on coherent strain-rate,
and $(iv)$ exertion of mixed stress on residual strain-rate. For conciseness of this paper, we abstain
from analysing all the single contributions of this mixed flux regarding their rotation angles,
and focus on the sum of all four contributions altogether.

\subsection{Multi-scale gradient (MSG) flux expansion}\label{section4_3}
As a final extension of the flux analysis, the locality between
strain-rate tensors on varying scales is analysed according to the
second-order MSG approach 
\citep{Eyink2006,Eyink2006b}.
For that, a second filtering operation is defined as
\begin{align}
u_i^{(b)}(\boldsymbol{x}) = \int G_{\ell_b}(\boldsymbol{r}) u_i(\boldsymbol{x}+\boldsymbol{r}) \diffunit^2 \boldsymbol{r}, \label{eq:multiscale_filtered_velocity}
\end{align}
where $G_{\ell_b}$ filters out contributions from all scales smaller than
$\ell_b = \lambda^{-b} \ell$, with a geometric factor $\lambda > 1$. 
This leads to the band-pass filtered velocity
\begin{align}
u_i^{[b]} &= \begin{cases}
      u_i^{(b)} - u_i^{(b-1)}, \quad b \geq 1, \\
      \overline{u}_i, \quad b = 0,
    \end{cases} \label{eq:bandpass_filtered_velocity}
\end{align}
representing contributions from a band of length scales 
between $\ell_b$ and $\ell_{b-1}$. 
The filtering operation leads to the multi-scale 
property of the MSG expanded cross-scale flux approach. 
The multi-gradient nature comes from a
Taylor expansion of the velocity increments
$\delta \boldsymbol{u}(\boldsymbol{r};\boldsymbol{x}) = \boldsymbol{u}(\boldsymbol{x}+\boldsymbol{r}) - 
\boldsymbol{u}(\boldsymbol{x})$ with separation vector $\boldsymbol{r}$.
The technical details for the derivation of the second-order MSG 
flux are outlined in appendix \ref{appendix1_3} and yield \citep[see][]{Eyink2006b}:
\begin{align}
 Z_{*}^{MSG} &= - \mathsfbi{S}^{(0)} : \boldsymbol{\tau}_{*}^{MSG} \nonumber \\
             &= Z_{*}^{MSG}(\mathsfbi{S}^{(0)} : \tilde{\mathsfbi{S}}^{[b]},\mathsfbi{S}^{(0)} : \mathsfbi{S}^{[b]}, (\nabla \omega^{[b]})^{T} \mathsfbi{S}^{(0)} (\nabla \omega^{[b]})) \nonumber \\
             &= \sum_{b=0}^{n_b} (Z_{SR}^{[b]} + Z_{DSR}^{[b]} + Z_{DSM}^{[b]} + Z_{VGS}^{[b]}) - Z_{FSF}^{(n_b)}. \label{eq:MSG_CSA_flux}
\end{align}
The parameter $b \in \mathbb{N}_0$ denotes the level of scale locality
of the respective MSG flux contributions $Z_{SR}^{[b]}$, $Z_{DSR}^{[b]}$,
$Z_{DSM}^{[b]}$ and $Z_{VGS}^{[b]}$, meaning that for low $b$-values 
contributions from strongly scale local interactions are measured, 
whereas contributions of non-local interactions are obtained for larger values.
The total number of filter bands is denoted as $n_b$.
The inner products between
tensors, as well as matrix vector products are expressible in polar
coordinates as
\begin{align}
 \mathsfbi{S}^{(0)} : \tilde{\mathsfbi{S}}^{[b]} &= -\sigma^{(0)} \sigma^{[b]} \sin(2 \delta \alpha^{[b]}), \\
 \mathsfbi{S}^{(0)} : \mathsfbi{S}^{[b]} &= \sigma^{(0)} \sigma^{[b]} \cos(2 \delta \alpha^{[b]}), \\
 (\nabla \omega^{[b]})^{T} \mathsfbi{S}^{(0)} (\nabla \omega^{[b]}) &= -\sigma^{(0)} \left| \nabla \omega^{[b]} \right|^2 \cos(2 \delta \beta^{[b]}),
\end{align}
where $\sigma^{(0)}$ and $\sigma^{[b]}$ are the positive eigenvalues
of the strain-rate tensors $\mathsfbi{S}^{(0)}$ and $\mathsfbi{S}^{[b]}$,
respectively, with $\alpha^{(0)}$ and $\alpha^{[b]}$ the angles
between their corresponding eigenvectors to a fixed orthogonal frame
of reference, and $\tilde{\mathsfbi{S}}^{[b]}$ the skew-strain-rate matrix
rotated counterclockwise by $\pi/4$ to the original strain matrix
$\mathsfbi{S}^{[b]}$. According to figure \ref{fig:rotation_angle_overview}
(b), $\delta \alpha^{[b]} = \alpha^{[b]} - \alpha^{(0)}$ is the
rotation angle between the large-scale tensor $\mathsfbi{S}^{(0)}$ and the
subfilter-scale tensors $\mathsfbi{S}^{[b]}$. Figure
\ref{fig:rotation_angle_overview} (c) shows $\delta \beta^{[b]}$,
which is the angle between the vorticity gradient vector $\nabla
\omega^{[b]}$ and the eigenvector of $\mathsfbi{S}^{(0)}$ corresponding to
the negative eigenvalue. The latter is equivalent to its contractile
direction.

The second-order MSG flux can be subdivided into four flux channels,
in which the investigation of the angles $\delta \alpha^{[b]}$ and
$\delta \beta^{[b]}$ directly illuminates the proposed vortex thinning
picture \citep{Eyink2006b,XiaoWanChenEyink2009}:
\begin{itemize}
\item The strain rotation (SR) $Z_{SR}^{[b]}$ is equivalent to the first-order
 MSG expansion 
 and relates to  the following physical
 picture: A small-scale vortex $\omega^{[b]}$ embedded in a
 large-scale strain-rate field $\mathsfbi{S}^{(0)}$, as illustrated in figure
 \ref{fig:rotation_angle_overview} (b), is stretched along the
 positive and compressed along the negative eigendirection of the
 strain. This leads to an elliptical shape inducing a shear layer and
 thus a small-scale strain rotated with $\delta \alpha^{[b]} = \pm
 \pi/4$ towards the large-scale strain, depending on the sign of the
 vorticity.
\item The differential strain rotation (DSR) $Z_{DSR}^{[b]}$ contains
 a Newtonian stress-strain relation of the form $\boldsymbol{\tau}^{[b]} =
 -\nu_T^{[b]} \mathsfbi{S}^{[b]}$, with negative eddy-viscositiy
 $\nu_T^{[b]}$. According to figure \ref{fig:rotation_angle_overview}
 (b), the elliptically-shaped vortex still possesses the same area,
 but the circumference increases leading to a loss of energy, due to
 Kelvin's theorem of the conservation of circulation $\Gamma = \oint
 \boldsymbol{u} \bcdot \diffunit \boldsymbol{s}$. As a consequence, the small-scale
 stress $\boldsymbol{\tau}^{[b]}$ exerts negative work on the large-scale
 strain $\mathsfbi{S}^{(0)}$ because of its parallel alignment to that
 strain. This results in an energy transfer towards larger length
 scales.
\item The vorticity gradient stretching (VGS) $Z_{VGS}^{[b]}$ is a measure of
 the elongation of vortex lines. The angle $\delta \beta^{[b]}$ between the
 vorticity gradient vector $\nabla \omega^{[b]}$ and the contractile
 direction of the large-scale strain $\mathsfbi{S}^{(0)}$ measures the
 alignment of the stretching direction to the vorticity
 isolines. Thus, a higher tendency for this alignment increases the
 rate of stretching parallel to the isolines, as depicted in figure
 \ref{fig:rotation_angle_overview} (c), leading to a thinning of the
 vortex.  
\item The differential strain magnification (DSM) $Z_{DSM}^{[b]}$
 contains, similar to the SR term, a \textit{skew-Newtonian}
 stress-strain relation with \textit{skew-eddy-viscositiy}
 $\gamma_T^{[b]}$. It measures the logarithmic rate of strain
 increase, when moving in the direction of increasing
 vorticity. According to \citet{XiaoWanChenEyink2009}, this term is
 generally expected to be smaller, as we can confirm in the results of
 section \ref{section6_2}.
\end{itemize}

Although the vortex thinning picture is not necessarily associated
with single coherent vortices, but rather with the whole vorticity
ensemble itself, we intend to measure the influence of coherent
regions and their residual backgrounds regarding this mechanism. This
is achieved by the following three-part decomposition:
\begin{align}
 & Z_{*}^{MSG} = \underbrace{- \mathsfbi{S}_c^{(0)} : \boldsymbol{\tau}_{*,c,c}^{MSG}}_{Z_{*,c}^{MSG}} \underbrace{- \mathsfbi{S}_r^{(0)} : \boldsymbol{\tau}_{*,r,r}^{MSG}}_{Z_{*,r}^{MSG}} \nonumber \\
 & \underbrace{- \mathsfbi{S}_c^{(0)} : \boldsymbol{\tau}_{*,r,r}^{MSG} - \mathsfbi{S}_r^{(0)} : \boldsymbol{\tau}_{*,c,c}^{MSG} - \mathsfbi{S}_c^{(0)} : (\boldsymbol{\tau}_{*,c,r}^{MSG} + \boldsymbol{\tau}_{*,r,c}^{MSG}) - \mathsfbi{S}_r^{(0)} : (\boldsymbol{\tau}_{*,c,r}^{MSG} + \boldsymbol{\tau}_{*,r,c}^{MSG})}_{Z_{*,cr}^{MSG}}, \label{eq:decomposed_MSG_CSA_flux}
\end{align}
with $Z_{*,c}^{MSG}$ and $Z_{*,r}^{MSG}$ the purely coherent and
residual second-order MSG flux expansions, respectively, and
$Z_{*,cr}^{MSG}$ the flux contribution originating from the mixed
interactions. 
For the reasons similar to those already given in section \ref{section4_2},
the same form of heterogeneous stresses, 
$\boldsymbol{\tau}_{*,c,r}^{MSG}$ and $\boldsymbol{\tau}_{*,r,c}^{MSG}$, appears in the mixed
MSG expanded flux. 
To limit the scope of this paper, we refrain from an in-depth analysis 
of the vortex thinning angles $\delta \alpha^{[b]}$ and $\delta \beta^{[b]}$
for the mixed MSG flux contribution.
However, the decomposition of the MSG expanded flux into purely coherent
and residual parts implies a decomposition of the different flux
channels as well
\begin{align}
 Z_{*,c/r}^{MSG} = \sum_{b=0}^{n_b} (Z_{SR,c/r}^{[b]} + Z_{DSR,c/r}^{[b]} + Z_{DSM,c/r}^{[b]} + Z_{VGS,c/r}^{[b]}) - Z_{FSF,c/r}^{(n_b)}. \label{eq:coherent_residual_MSG_CSA_flux}
\end{align}
This leads to the analysis of different angles between strain-rate
tensors and vorticity gradient vectors for varying scale localities, 
set by $b$, originating from coherent and residual components
\begin{align}
 \boldsymbol{S}^{(0)}_c : \tilde{\boldsymbol{S}}^{[b]}_c &= -\sigma^{(0)}_c \sigma^{[b]}_c \sin(2 \delta \alpha^{[b]}_c), \label{eq:decomposed_polar_coordinated_first} \\
 \mathsfbi{S}^{(0)}_r : \tilde{\mathsfbi{S}}^{[b]}_r &= -\sigma^{(0)}_r \sigma^{[b]}_r \sin(2 \delta \alpha^{[b]}_r), \\
 \mathsfbi{S}^{(0)}_c : \mathsfbi{S}^{[b]}_c &= \sigma^{(0)}_c \sigma^{[b]}_c \cos(2 \delta \alpha^{[b]}_c), \\
 \mathsfbi{S}^{(0)}_r : \mathsfbi{S}^{[b]}_r &= \sigma^{(0)}_r \sigma^{[b]}_r \cos(2 \delta \alpha^{[b]}_r), \\
 (\nabla \omega^{[b]}_c)^{T} \mathsfbi{S}^{(0)}_c (\nabla \omega^{[b]}_c) &= -\sigma^{(0)}_c \left| \nabla \omega^{[b]}_c \right|^2 \cos(2 \delta \beta^{[b]}_c), \\
 (\nabla \omega^{[b]}_r)^{T} \mathsfbi{S}^{(0)}_r (\nabla \omega^{[b]}_r) &= -\sigma^{(0)}_r \left| \nabla \omega^{[b]}_r \right|^2 \cos(2 \delta \beta^{[b]}_r). \label{eq:decomposed_polar_coordinated_last}
\end{align}
The variables are interpreted in the same fashion as above 
for the total field but now with 
respect to the coherent (index $c$) and residual (index $r$)
contributions. We present an analysis of the thinning
effects in section \ref{section6_2},
for which $Z_{SR,c/r}^{[b]}$, $Z_{DSR,c/r}^{[b]}$, $Z_{VGS,c/r}^{[b]}$
and their corresponding angles $\delta \alpha^{[b]}_{c/r}$, $\delta \beta^{[b]}_{c/r}$
are the relevant quantities measuring the thinning tendencies
of purely coherent and residual parts, respectively.

\section{Numerical methods and parameters}\label{section5}
Equation \eqref{eq:navier_stokes1} and \eqref{eq:navier_stokes2} are solved in Fourier space using the equivalent and numerically
more favourable vorticity representation. 
The differential equation includes a small-scale
forcing term, $\hat{f}_\omega$, and a large-scale damping function, 
$-\hat{d}_\omega\hat{\omega}$, yielding
\begin{align}
 \partial_t \hat{\omega} + \widehat{\left[ (\boldsymbol{u} \bcdot \nabla) \omega \right]} = \nu k^2 \hat{\omega} + \hat{f}_{\omega} - \hat{d}_{\omega} \hat{\omega}. \label{eq:navier_stokes_numerical_implementation}
\end{align}
It is solved by a pseudospectral approach, with a second-order
trapezoidal Leapfrog time integration scheme, and a $2/3$-dealiasing
method
\citep{CanutoHussainiQuarteroniZang_SpectralMethodsInFluidDynamics_1988}. The
forcing components of the velocity field $\hat{f}_{u,x}$ and
$\hat{f}_{u,y}$ are drawn from Gaussian normal distributions and they are
afterwards projected onto the solenoidal components $\hat{f}_{u,j} =
\left( \delta_{ij}-k_i k_j/k^2 \right) \hat{f}_{u,i}$ to satisfy the
incompressibility condition. The forcing term is then constructed as
$\hat{f}_{\omega}(\boldsymbol{k}) = \imagunit \left( k_x \hat{f}_{u,y}(\boldsymbol{k})
- k_y \hat{f}_{u,x}(\boldsymbol{k}) \right)$ and applied at a wavenumber of
$k_f=200$. In order to avoid the accumulation of energy at large
scales due to the inverse cascade, a large-scale linear damping term
with a Gaussian damping factor $\hat{d}_{\omega}(\boldsymbol{k}) =
\alpha_{\omega} \exp(-(k-k_{0,\omega})^2/(2 \sigma_{\omega}^2))$ is
employed. The parameters for the large-scale friction factor
$\alpha_{\omega}$, the center of the Gaussian damping profile
$k_{0,\omega}$ and its variance $\sigma^2_{\omega}$ are given in table
\ref{tab:simulation_parameters_inverse_kinetic_energy_cascade_hydro_runs}
in appendix \ref{appendix2}. We solve the system at a resolution of
$4096^2$ in the square periodic domain $2 \pi \times 2 \pi$.

\begin{figure}
\centering
\captionsetup{width=\textwidth}
\includegraphics[width=\textwidth]{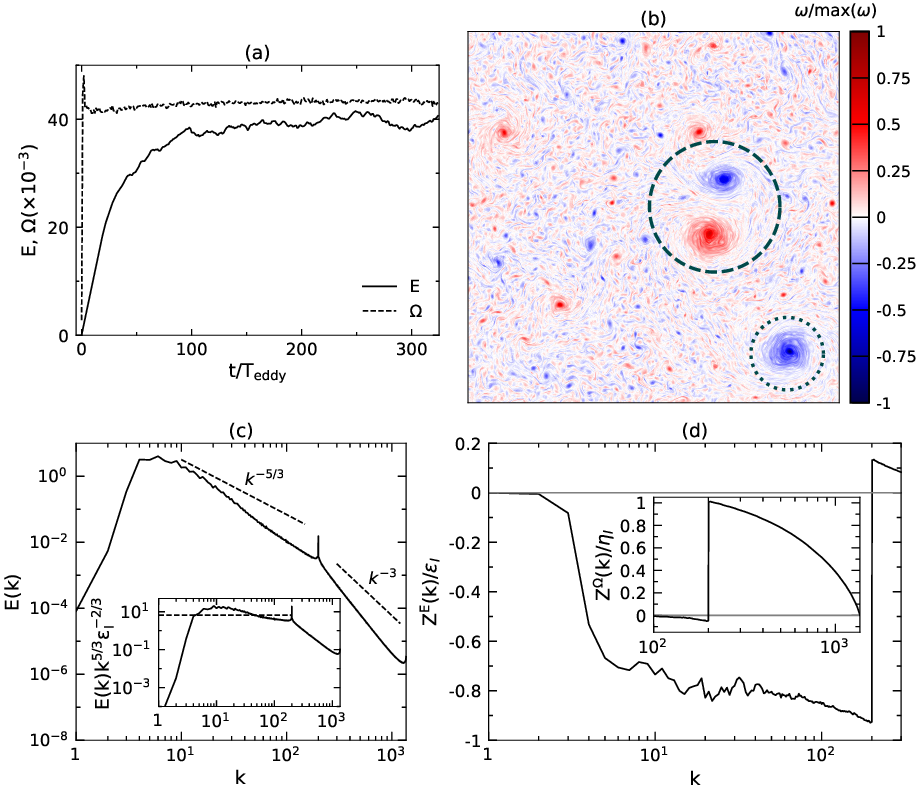}
\caption{Flow observables of run1 (see table 
\ref{tab:simulation_parameters_inverse_kinetic_energy_cascade_hydro_runs} 
in appendix \ref{appendix2}). (a): Time evolution of the kinetic energy $E$ and 
enstrophy $\Omega$. The enstrophy is divided by $10^3$ for better visualisation. (b): 
Vorticity $\omega$ from a $1024^2$ region, which is $6.25\%$ of the total physical domain. 
The colorbar is normalised to $\omega/\max{|\omega|}$ for better visualisation of emerging 
structures. The upper circle highlights a vortex pair and the lower circle marks a single 
large-scale vortex. Physical quantities for both of these regions are further analysed in 
figure \ref{fig:various_physical_quantities_run2_4}. (c): Kinetic energy spectrum $E(k)$ with 
inset showing the compensated spectrum $E(k) k^{5/3} \epsilon_I^{-2/3}$, where the black 
dashed line indicates a value of $C_E = 6.69$ predicted by the \textit{test-field model} 
(TFM) closure of \citet{Kraichnan1971}. (d): Normalised cross-scale kinetic energy flux 
$Z^E(k)/\epsilon_I$, where the inset shows the normalised enstrophy flux $Z^{\Omega}(k)/\eta_I$.}%
\label{fig:time_evolution_vorticity_spectra_fluxes_run2}%
\end{figure}

The large-eddy turnover time is estimated as $T_{eddy} =
L_{int}/u_{rms}$, with $L_{int}=\int k^{-1} E(k) dk / \int E(k) dk$
the integral length scale and $u_{rms}$ the root-mean square velocity,
which is also used for the integration time of the passive tracers in
the FTLE/LCS calculation in equation \eqref{eq:FTLE}. Our results are
taken after reaching a statistically stationary state, as confirmed in
figure \ref{fig:time_evolution_vorticity_spectra_fluxes_run2} (a). They 
are averaged over $100$ snapshots equidistantly distributed over
roughly $20 T_{eddy}$.

A discussion of the chosen values of the energy injection rate
$\epsilon_I$ and the general system parameters for the present
numerical setup can be found in appendix \ref{appendix2}, which is related to
the characteristics of structure formation, the kinetic energy
spectrum and the cross-scale kinetic energy flux. There, we conclude
that run1 (table
\ref{tab:simulation_parameters_inverse_kinetic_energy_cascade_hydro_runs}
in appendix \ref{appendix2}) is the best choice for the purpose of our
present study. The spatial vorticity distribution in figure
\ref{fig:time_evolution_vorticity_spectra_fluxes_run2} (b) exhibits a
clearly developed population of visually distinguishible vortices or
coherent structures.
The kinetic energy spectrum in figure
\ref{fig:time_evolution_vorticity_spectra_fluxes_run2} (c) deviates
from the theoretically expected $k^{-5/3}$ scaling due to
finite-size effects discussed in appendix \ref{appendix2},
but we deem it to be more adequate for the subsequent analysis due to a 
more clearly discernible structuring of the flow. The cross-scale kinetic
energy and enstrophy fluxes shown in figure
\ref{fig:time_evolution_vorticity_spectra_fluxes_run2} (d) possess 
sufficiently extended ranges of inverse and
direct spectral transfer. This facilitates the cross-scale flux
decompositions in sections \ref{section6_1} and \ref{section6_2}.

\subsection{Structure detection}\label{section5_1}
\begin{figure}
\centering
\captionsetup{width=\textwidth}
\includegraphics[width=\textwidth]{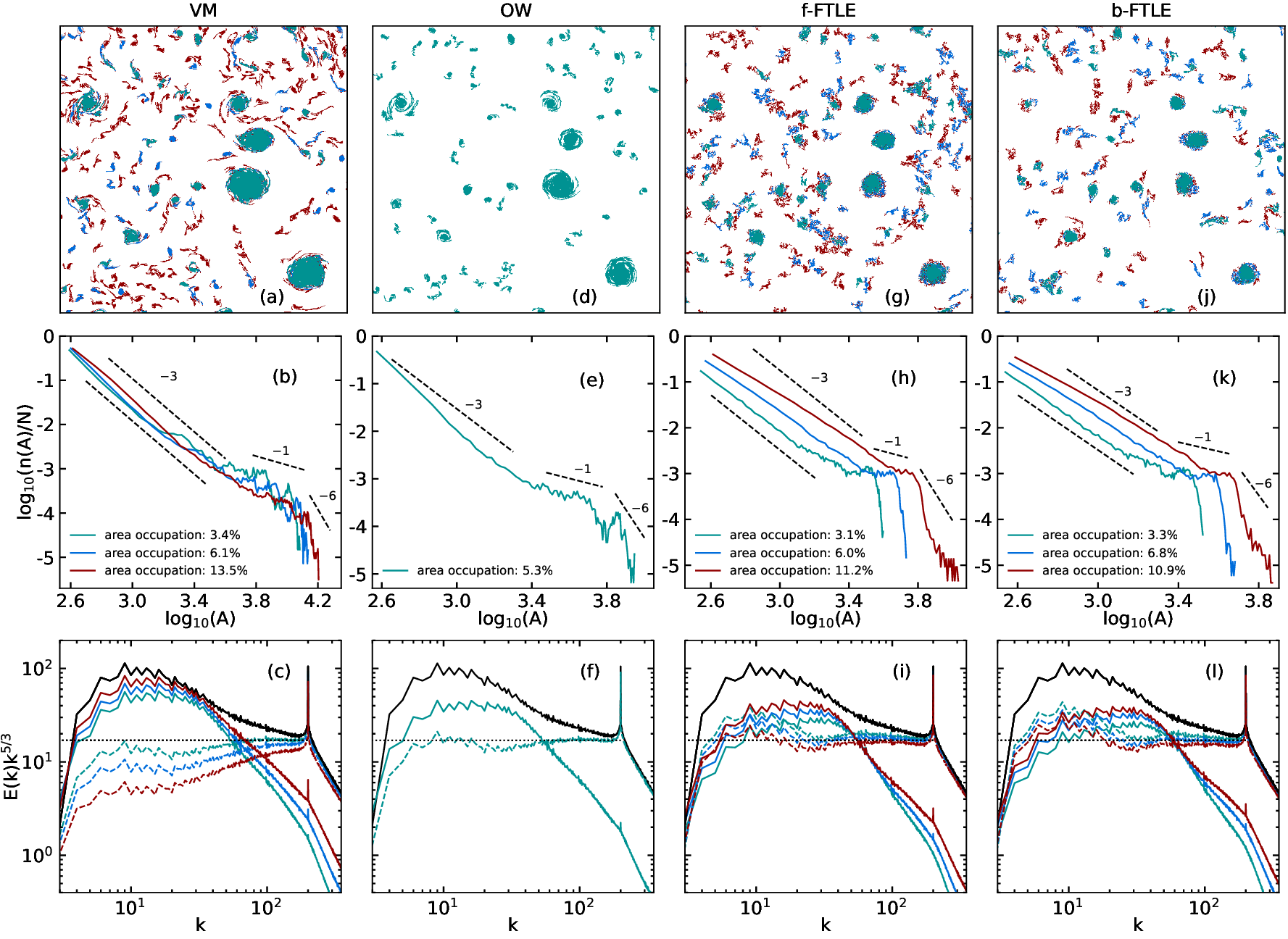}
\caption{Coherent structures, number density and kinetic energy spectra obtained by 
the VM (a)-(c), OW criterion (d)-(f), f-FTLE 
(g)-(i) and b-FTLE (j)-(l), respectively. {The coherent areas are 
color-coded matching the normalised number densities $n(A)/N$ and the decomposed 
energy spectra, with $N$ the total number of detected structures. Color-code: $(i)$ 
lowest area occupation: green regions corresponding to the green lines of the number 
densitiy and decomposed energy spectra plots, $(ii)$ intermediate area occupation: 
green+blue regions corresponding to the blue lines and $(iii)$ largest area occupation: 
green+blue+red regions corresponding to the red lines.} The OW criterion has 
only one permitted threshold, hence only a single area occupation is shown. All the 
energy spectra are compensated by $k^{-5/3}$, with the black solid line indicating the 
total energy spectrum $E(k)$, and the colored solid and dashed lines denoting the 
coherent $E_{c}(k)$ and residual $E_{r}(k)$ spectra, respectively. The black dotted 
horizontal line visualises deviations from the $k^{-5/3}$ scaling.}
\label{fig:number_density_vorticity_threshold_fwdFTLE_bwdFTLE_okuboweiss_varying_area_occupation_run2_3}
\end{figure}

As already mentioned in the previous sections \ref{section3_2} and
\ref{section3_3}, the threshold choice for the VM,
the f-FTLE and the b-FTLE criterion to sample coherent regions
from the vorticity distribution is not straightforward. Therefore, the
threshold is chosen such that the three-subregime structure of the coherent vortex number density, $n(A)$,
in equation
\eqref{eq:three_part_number_density} is realized most clearly.
The system studied by \citet{BurgessScott2017,BurgessScott2018} 
assumed stationarity by
imposing an integral length scale far below the largest length scales
of the system domain. On the contrary, our system is in a
statistically stationary state with constant kinetic energy $E$ and
enstrophy $\Omega$ according to figure
\ref{fig:time_evolution_vorticity_spectra_fluxes_run2} (a). Therefore,
we analyse the scaling sensitivity of the number density $n(A)$ only
in dependence of the coherent area $A$ without the time $t$, as
presented in figures
\ref{fig:number_density_vorticity_threshold_fwdFTLE_bwdFTLE_okuboweiss_varying_area_occupation_run2_3}
(b), (e), (h) and (k).

Our comparison of coherence specifications and detection methods begins with the VM criterion.
This approach is technically closest to the detection method chosen in \cite{BurgessScott2017,BurgessScott2018}
and thus expected and observed to yield the best agreement with the findings published there 
and with the corresponding asymptotic scaling laws of the vortex number density, $n(A)$.
We vary the VM threshold introduced in section \ref{section3_2} 
with three different values
$\epsilon_{thr}=1.1 \omega_{rms}$, $0.9 \omega_{rms}$, $0.7 \omega_{rms}$, 
for which the total areas occupied by the coherent regions in figure
\ref{fig:number_density_vorticity_threshold_fwdFTLE_bwdFTLE_okuboweiss_varying_area_occupation_run2_3}
(a) amount to $3.4\%$, $6.1\%$, $13.5\%$, respectively. Filamentary structures
occur with increasing area occupation extending the
thermal bath regime in figure
\ref{fig:number_density_vorticity_threshold_fwdFTLE_bwdFTLE_okuboweiss_varying_area_occupation_run2_3}
(b). The number density approximately exhibits the phenomenologically proposed asymptotic 
$A^{-3}$ scaling for the highest area occupation.
The approximate power-law deteriorates
as the threshold is raised and the detected set comprises of fewer structures.
As the VM criterion explicitly 
detects regions of high vorticity, 
the most energetic regions of the flow are included
in the coherent part, which is evident from the decomposed kinetic energy spectra 
in figure \ref{fig:number_density_vorticity_threshold_fwdFTLE_bwdFTLE_okuboweiss_varying_area_occupation_run2_3}
(c). Most of the energy is concentrated in $E_c(k)$,
whereas the energy in the residual spectrum $E_r(k)$ 
decreases for larger length scales with scalings
which become flatter than $k^{-5/3}$ for increasing
area occupations. The clearly discernible intermediate region 
exhibits an increased fluctuation level for lower
thresholds since larger and thus fewer structures tend to be detected. 
This is consistent with larger deviations of the measured scaling
exponent from the predicted value than in the case of the thermal bath 
as seen in table \ref{tab:power_law_scalings_vortex_number_density}.

In contrast to the VM method, the OW criterion only permits one
possible threshold $\epsilon_{thr}=0$.
The detected coherent regions in figure
\ref{fig:number_density_vorticity_threshold_fwdFTLE_bwdFTLE_okuboweiss_varying_area_occupation_run2_3}
(d) amount to an area occupation of $5.3\%$.
The thermal bath range
obtained via the OW method drops steeper with increasing area $A$ than for the VM technique 
(cf. table \ref{tab:power_law_scalings_vortex_number_density}).
This indicates a preference of the OW criterion for intense vortices at the cost of lesser 
vortical structures in the interval $A_f \leq A < A_-$. 
The
intermediate range of the number density $n(A)$ in figure
\ref{fig:number_density_vorticity_threshold_fwdFTLE_bwdFTLE_okuboweiss_varying_area_occupation_run2_3}
(e) has the best fulfillment of the $A^{-1}$ scaling compared to the
other methods
according to table \ref{tab:power_law_scalings_vortex_number_density}.
This is not surprising, since the OW 
criterion favours circularly shaped structures, 
which are also detected by the modified vorticity 
threshold criterion used by
\citet{BurgessScott2017,BurgessScott2018}. 
The residual energy spectrum $E_r(k)$ in figure
\ref{fig:number_density_vorticity_threshold_fwdFTLE_bwdFTLE_okuboweiss_varying_area_occupation_run2_3}
(f) possesses a scaling closer to $k^{-5/3}$ in the inertial range 
from $k \approx 10-200$ versus the total spectrum $E(k)$. 
In comparison, the coherent spectrum $E_c(k)$ has a much 
shorter $k^{-5/3}$ scaling range from $k \approx 10-30$, 
which becomes shallower for increasing wavenumbers.
This indicates that the lacking energetic self-similarity of largest-scale coherent 
structures detected by the OW criterion pollutes the scaling of the 
theoretically expected KLB spectrum of total energy (see appendix \ref{appendix2}). 
The residual energy whose scaling is not suffering from this specific finite-size 
effect of the numerical simulation exhibits good agreement with KLB expectations, see also, e.g.,
\citep{Borue1994,Scott2007,Vallgren2011,BurgessScott2018}.

With regard to the sensitivity of the LCS detection, presented in section \ref{section3_3}, 
we set the thresholds of the f-FTLE 
sampling scheme to $\epsilon_{thr} = 0.87
\Lambda_{t_0,rms}^{t_0+T_{eddy}}$, $0.905
\Lambda_{t_0,rms}^{t_0+T_{eddy}}$, $0.94
\Lambda_{t_0,rms}^{t_0+T_{eddy}}$, which amount to coherent area
occupations of $3.1\%$, $6.0\%$, $11.2\%$, respectively.
The thresholds for the
b-FTLE are set to
$\epsilon_{thr} = 0.88
\Lambda_{t_0,rms}^{t_0-T_{eddy}}$, $0.915
\Lambda_{t_0,rms}^{t_0-T_{eddy}}$, $0.94
\Lambda_{t_0,rms}^{t_0-T_{eddy}}$, resulting in coherent area
occupations of $3.3\%$, $6.8\%$, $10.9\%$, respectively.
The \textit{crinkly}-shaped
structures occuring in the domain, as shown in figures
\ref{fig:number_density_vorticity_threshold_fwdFTLE_bwdFTLE_okuboweiss_varying_area_occupation_run2_3}
(g) and (j), lead to the flattening of the thermal bath regime
and a simultaneous steepening of the intermediate range
with increasing area occupation, as depicted in figures
\ref{fig:number_density_vorticity_threshold_fwdFTLE_bwdFTLE_okuboweiss_varying_area_occupation_run2_3}
(h) and (k). 
Note that the different structural shape also
results in  a more extended thermal bath regime and a diminishing intermediate range 
with increasing area occupation in contrast to the OW 
and VM criterion.
Comparing f-FTLE and b-FTLE fields, 
similar larger-sized structures are detected by both fields, whereas 
smaller-sized structures are detected at differing positions.
This is not surprising, since forward- and backward-in-time LCSs, are associated with
different fluid dynamics, i.e. repelling and attracting manifolds in a 
dynamical systems sense, respectively \citep{HallerYuan2000,Haller2015}.
The residual energy spectra $E_r(k)$ in figures
\ref{fig:number_density_vorticity_threshold_fwdFTLE_bwdFTLE_okuboweiss_varying_area_occupation_run2_3}
(i) and (l) are closer to a $k^{-5/3}$ scaling throughout the entire inertial range
compared to the total spectrum $E(k)$, indicating a pollution of the KLB spectrum 
by the coherent structures similar to the results of the OW criterion.

The large-scale damping required for
the achieving stationarity impacts the
apparition of large-scale structures as discussed in appendix
\ref{appendix2}. This leads to the anomalous and 
partially polluted power-laws of the front regime in figures
\ref{fig:number_density_vorticity_threshold_fwdFTLE_bwdFTLE_okuboweiss_varying_area_occupation_run2_3}
(b), (e), (h) and (k) compared to the phenomenologically expected exponent of
$A^{-6}$ \citep{BurgessScott2017,BurgessScott2018}, where a
large-scale damping mechanism has not been employed. Another reason for the deviation is
the resolution of $8192^2$ in the conducted DNS of
\citet{BurgessScott2017,BurgessScott2018}, which is higher compared to
our present study. Nevertheless, the overall trend of
the three-part number density $n(A)$ is satisfied in all of
these definitions. Moreover, independent of the definition, coherent
structures tend to distort the theoretically predicted KLB scaling as
illustrated in figures
\ref{fig:number_density_vorticity_threshold_fwdFTLE_bwdFTLE_okuboweiss_varying_area_occupation_run2_3}
(c), (f), (i) and (l). 

In summary and in comparison to the VM detection technique the OW and the LCS approaches tend to detect similar large-scale
coherence features of the flow, in spite of their distinct underlying coherence specifications. The different characterizations of
coherence lead to a reduced sensitivity for small-scale turbulent structures in the case of OW, while the physically most involved LCS method
tends to detect a surplus of small-scale structures compared to the most simple VM specification.
We proceed by setting the respective detection parameters such that the detection signatures as shown in figures 
\ref{fig:number_density_vorticity_threshold_fwdFTLE_bwdFTLE_okuboweiss_varying_area_occupation_run2_3} (b), (e), (h) and (k) 
become most similar to each other,
i.e. an area occupation of 6.1\% for VM, of 6.0\% for f-FTLE, and of 6.8\% for b-FTLE.
The resulting scalings in table \ref{tab:power_law_scalings_vortex_number_density}
reflect a rough overall agreement with the
three expected scaling regimes of the number density $n(A)$.
Small adjustments of these thresholds do not result into
qualitatively significant differences regarding 
the inverse cascade analysis, which is 
conducted in the next section.

\begin{table}
\captionsetup{width=\textwidth}
\setlength{\tabcolsep}{1.5pt}
\begin{center}
\begin{tabular}{l|c|c|c|c|c|c|c}
Method & $\log_{10}(A_f)$ & $\log_{10}(A_-)$ & $\log_{10}(A_+)$ & $\log_{10}(A_{\max})$ & $(i)$ &  $(ii)$ & $(iii)$ \\[3pt]
\hline
VM & $2.6$ & $3.44$ & $3.94$ & $4.14$ & $-2.92 \pm 0.09$ & $-1.38 \pm 0.11$ & $-5.15 \pm 0.86$ \\
OW & $2.57$ & $3.42$ & $3.72$ & $3.95$ & $-3.41 \pm 0.07$ & $-1.01 \pm 0.13$ & $-4.63 \pm 0.66$ \\
f-FTLE & $2.57$ & $3.53$ & $3.66$ & $3.73$ & $-2.63 \pm 0.01$ & $0.11 \pm 0.31$ & $-24.45 \pm 2.24$ \\
b-FTLE & $2.56$ & $3.45$ & $3.58$ & $3.68$ & $-2.8 \pm 0.02$ & $0.08 \pm 0.31$ & $-23.79 \pm 1.68$ \\
\end{tabular}
\caption{Logarithmic power-law scalings of the three-part vortex number densities 
$n(A)$ corresponding to structures detected by the VM 
($\epsilon_{thr}=0.9 \omega_{rms}$, area occupation: $6.1\%$), OW criterion, f-FTLE 
($\epsilon_{thr} = 0.905 \Lambda_{t_0,rms}^{t_0+T_{eddy}}$, area occupation: 
$6.0\%$) and b-FTLE ($\epsilon_{thr} = 0.915 \Lambda_{t_0,rms}^{t_0-T_{eddy}}$, 
area occupation: $6.8\%$), respectively.
The scaling ranges are denoted as: $(i)$ 
thermal bath $A_f \leq A < A_-$, $(ii)$ intermediate $A_- < A < A_+$, and $(iii)$ 
front $A_+ < A \leq A_{max}$. Scaling exponents based on linear regression are given together with the respective standard deviation. The transitional areas $A_-$ and $A_+$ are estimated 
from figures 
\ref{fig:number_density_vorticity_threshold_fwdFTLE_bwdFTLE_okuboweiss_varying_area_occupation_run2_3} 
(b), (e), (h) and (k). For simplicity, the forcing-scale area $A_f$ and the maximum 
vortex area $A_{\max}$ are chosen as the first and last datapoints of the respective 
abscissae of $n(A)$.}
\label{tab:power_law_scalings_vortex_number_density}
\end{center}
\end{table}

\section{Relation of coherent structures to the inverse cascade process}\label{section6}
Coherent features of the vorticity field $\omega$ (shown in  figure
\ref{fig:time_evolution_vorticity_spectra_fluxes_run2} (b)) contain most of its kinetic 
energy, as inferred from the spatial distribution from the
absolute velocity $\left| \boldsymbol{u} \right|$ in figure
\ref{fig:various_physical_quantities_run2_4} (a), with the largest fraction
located in the vicinity of the vortex core. For the vortex pair
in figure \ref{fig:time_evolution_vorticity_spectra_fluxes_run2} (b),
the energy increases towards their separatrix where the maximum value is reached.

The f-FTLE and b-FTLE fields,
$\Lambda_{t_0}^{t_0+T_{eddy}}$ and $\Lambda_{t_0}^{t_0-T_{eddy}}$, are
illustrated in figure \ref{fig:various_physical_quantities_run2_4} (c)
and (d), respectively, where high FTLE values are potential candidates
for LCSs.
The LCS method considered here detects coherent vortices by the characteristic patterns of two-particle dispersion dynamics
perpendicular to the identified material lines that are shown in the figure.
From this perspective, the approach senses the imprint of a coherent structure on the surrounding flow rather than detecting
specific differential (OW) or amplitude markers (VM) associated with coherence. LCS can thus be regarded as complementary to the two other
methods considered here.
The LCS based on the f-FTLE field displays more pronounced small-scale fluctuations perpendicular to
the respective material line as 
compared to the b-FTLE field.
This reflects the different repelling and attracting 
dynamics expected along forward- and backward-in-time
LCSs, while the difference between the Lagrangian scheme
used for the f-FTLE and the semi-Lagrangian approach employed
for the b-FTLE (see appendix \ref{appendix1_2}) can play a role
here as well.
There is a strong visual correlation between ridges in
both FTLE fields with the boundaries of vortices observed in the
vorticity field. This corroborates the above choice of sampling the
vorticity distribution with the f-FTLE and b-FTLE according to
equations \eqref{eq:decomposition_coherent} and \eqref{eq:decomposition_residual}.

\subsection{Cross-scale flux efficiency}\label{section6_1}
\begin{figure}
\centering
\captionsetup{width=\textwidth}
\includegraphics[scale=1.5]{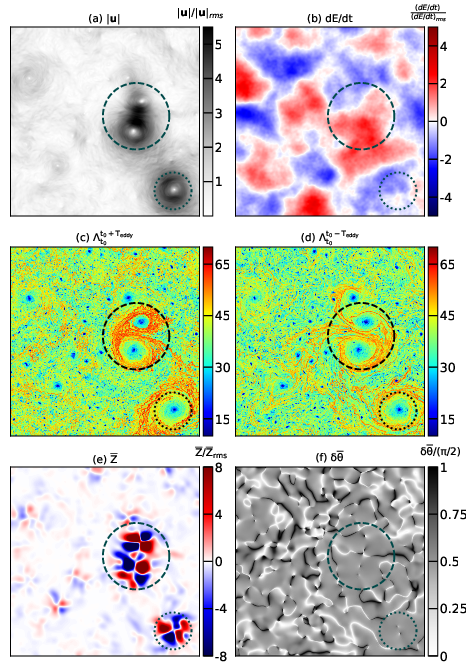}
\caption{Various physical quantities of the same region as shown in figure 
\ref{fig:time_evolution_vorticity_spectra_fluxes_run2} (b). (a): Absolute 
velocity $\left| \boldsymbol{u} \right|$. (b): Energy rate $dE/dt$. 
(c): f-FTLE $\Lambda_{t_0}^{t_0+T_{eddy}}$. (d): b-FTLE $\Lambda_{t_0}^{t_0-T_{eddy}}$. 
(e): Cross-scale kinetic energy flux 
$\overline{Z}$ obtained from a smooth Gaussian filter with a filtering wavenumber 
of $k=60$. (f): Angle between strain-rate and subgrid stress tensor 
$\delta \overline{\theta}$ obtained from the same filter as in (e).}
\label{fig:various_physical_quantities_run2_4}
\end{figure}

When aiming at establishing a link between the nonlinear cross-scale flux and the detected structures in configuration space,
the temporal derivative of energy, $dE/dt$, see 
figure \ref{fig:various_physical_quantities_run2_4} (b), is only of limited value as it does not yield clear localized 
signatures correlated with coherent vortices. 
In contrast, the spatial distribution of the cross-scale energy flux 
$\overline{Z}$ exhibits intense quadrupolar structures 
in coherent regions reflecting their high level of symmetry. This is shown in figure \ref{fig:various_physical_quantities_run2_4} (e)
and has been observed by \citet{XiaoWanChenEyink2009} and \citet{LiaoOuellette2013} as well. 
Thus, coherent regions lead to large local contributions to the cross-scale flux, but do not 
necessarily generate significant contributions to the spatially averaged net inverse flux 
$\langle \overline{Z} \rangle$.

\begin{figure}
\centering
\captionsetup{width=\textwidth}
\includegraphics[width=\textwidth]{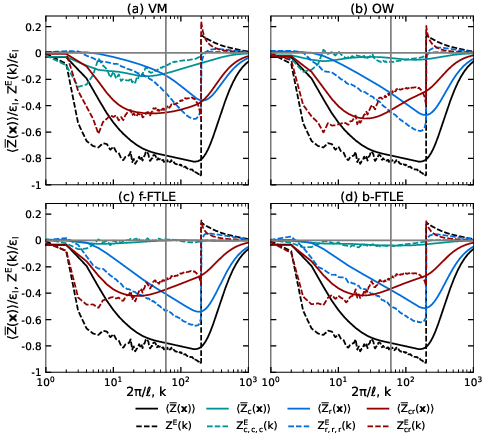}
\caption{Comparison of cross-scale kinetic energy fluxes of varying coherence definitions, 
obtained from the spectral formulation $Z^E(k)$ (dashed lines) and the spatially averaged 
formulation $\langle \overline{Z}(\boldsymbol{x}) \rangle$ (solid lines) obtained by 
Gaussian filtering for varying filter length scales $\ell$. The grey vertical line indicates 
the wavenumber $k=60$, at which the subsequent analysis of the spatial cross-scale flux 
distribution is conducted.}
\label{fig:comparison_decomposed_smooth_sharp_scale_energy_flux_vorticity_threshold_fwdFTLE_bkwdFTLE_okuboweiss_run2_4}
\end{figure}

The cross-scale energy flux is decomposed into its coherent, residual, 
and mixed contributions, for the spectral cross-scale fluxes 
$Z_{c,c,c}^E(k)$, $Z_{r,r,r}^E(k)$, $Z_{cr}^E(k)$ 
(equation \eqref{eq:decomposed_spectral_scale_flux}) 
and the spatially averaged cross-scale flux distributions in configuration space 
$\langle \overline{Z}_c(\boldsymbol{x}) \rangle$, 
$\langle \overline{Z}_r(\boldsymbol{x}) \rangle$, 
$\langle \overline{Z}_{cr}(\boldsymbol{x}) \rangle$ 
(equation \eqref{eq:decomposed_spatial_scale_flux}), respectively. 
Both cross-scale flux measurements are shown in figure 
\ref{fig:comparison_decomposed_smooth_sharp_scale_energy_flux_vorticity_threshold_fwdFTLE_bkwdFTLE_okuboweiss_run2_4}, 
because the $Z^E(k)$ representation is usually favored for quantifying the 
existence of turbulent cascade activity, whereas the $\overline{Z}(\boldsymbol{x})$ 
representation is commonly used to reveal the spatial structure of the flux. 
We observe that the fluxes of all coherence definitions are qualitatively 
similar, although the OW criterion in figure 
\ref{fig:comparison_decomposed_smooth_sharp_scale_energy_flux_vorticity_threshold_fwdFTLE_bkwdFTLE_okuboweiss_run2_4} 
(b), and the f-/b-FTLE in figures 
\ref{fig:comparison_decomposed_smooth_sharp_scale_energy_flux_vorticity_threshold_fwdFTLE_bkwdFTLE_okuboweiss_run2_4} 
(c) and (d), respectively, possess higher quantitative similarity in contrast to the VM scheme. 
For the latter, the coherent flux $Z_{c,c,c}^E(k)$ is slightly 
higher for all scales as shown in figure 
\ref{fig:comparison_decomposed_smooth_sharp_scale_energy_flux_vorticity_threshold_fwdFTLE_bkwdFTLE_okuboweiss_run2_4} 
(a) which shows the limits of the applied simple gauge criterion. This is because the VM favourably extracts higher-valued 
vorticity regions compared to the other detection schemes, which are attributed 
to larger-sized structures according to the coherent spectrum $E_c(k)$ in figure 
\ref{fig:number_density_vorticity_threshold_fwdFTLE_bwdFTLE_okuboweiss_varying_area_occupation_run2_3} 
(c). Therefore, the highest coherent flux contributions are rather in the 
lower wavenumber range, with a decreasing contribution towards higher wavenumbers.
The structures detected by the f-FTLEs and b-FTLEs have 
vanishing coherent flux contributions, 
which are close to zero for the entire inertial range.
This is in agreement with both 
forward- and backward-in-time LCSs, 
having the tendency to collectively inhibit the energy transfer among scales 
\citep{KelleyAllshouseOuellette2013}. The circularly-shaped structures identified 
by the OW criterion also exhibit the same behavior of inhibiting the 
cross-scale flux. These observations contribute to an alternative definition 
of coherence in a turbulence context in the sense that energy within 
these structures tends to remain rather closely at their given length scales 
without cascading across scales.

Generally, the merging of coherent vortices has been one of the most appealing 
physical mechanisms for the inverse cascade for quite some time. However, 
independent of the definition and as shown above, the coherent part of 
the flux $Z_{c,c,c}^E(k)$ has an overall low negative contribution 
throughout the inertial range. Therefore, merging effects of coherent structures have 
a weak influence to the overall inverse cascade, which also has been pointed out 
by \citet{XiaoWanChenEyink2009}. 

The residual flux $Z_{r,r,r}^E(k)$ remains negative 
throughout the inertial range as well, 
peaking at small-scales, 
close to the forcing wavenumbers, and decaying roughly
logarithmically towards larger-scales.
This is due to the contribution of detected smaller-sized structures, according to the 
residual spectra $E_r(k)$ already shown in 
figures \ref{fig:number_density_vorticity_threshold_fwdFTLE_bwdFTLE_okuboweiss_varying_area_occupation_run2_3} 
(c), (f), (i) and (l) above. 
We propose in section \ref{section6_2} that the negative contribution 
of the residual flux is attributed to a stronger impact of the thinning mechanism on smaller scales. 

Furthermore, a substantial amount of the net negative flux on each length scale 
originates from the mixed coherent-residual interactions $Z_{cr}^E(k)$, 
which roughly stays at a constant level throughout the inverse flux region.
We abstain from ascribing this flux contribution to more specific physical 
dynamics as this flux results from the complex interplay between coherent/residual/mixed stresses 
and coherent/residual strain-rates
as already mentioned in section \ref{section4_2}.

\begin{figure}
\centering
\captionsetup{width=\textwidth}
\includegraphics[width=\textwidth]{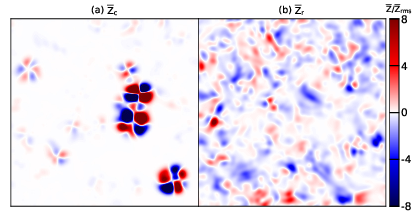}
\caption{Reconstructed spatial cross-scale kinetic energy flux distributions of the (a) 
purely coherent $\overline{Z}_c(\boldsymbol{x})$ and (b) purely residual 
$\overline{Z}_r(\boldsymbol{x})$ parts obtained by the VM criterion 
for a filtering wavenumber of $k=60$, with each normalised by its root-mean square value 
for visualisation purposes. The absolute values of $\overline{Z}_c$ are much higher 
compared to $\overline{Z}_r$ with a ratio 
$\frac{\max(|\overline{Z}_c|)}{\max(|\overline{Z}_r|)} \approx 44$.}
\label{fig:coherent_residual_spatial_scale_flux_distribution_vorticity_threshold_k_60_run2_2}
\end{figure}

Because the Fourier cross-scale flux is determined via spatial integration, an investigation 
in configuration space
is reasonable for gaining further insight. 
The coherent part $\overline{Z}_c(\boldsymbol{x})$ reconstructed 
from $\omega_c$ mostly consists of highly ordered quadrupolar structures 
and is exemplarily illustrated for the VM criterion in 
figure \ref{fig:coherent_residual_spatial_scale_flux_distribution_vorticity_threshold_k_60_run2_2} (a).  
Compared to that, the $\omega_r$-reconstructed residual part $\overline{Z}_r(\boldsymbol{x})$ in 
figure \ref{fig:coherent_residual_spatial_scale_flux_distribution_vorticity_threshold_k_60_run2_2} (b) 
has rather complex and unordered spatial features. 
Similar spatial characteristics for coherent and residual parts 
are obtained for the other coherence-detection methods. 
Also, qualitatively similar findings are gained for filtering lengths below 
the damping-dominated and above the forcing-dominated length scales, 
hence we employ a filtering wavenumber of $k=60$ for the remaining analysis. 

\begin{figure}
\centering
\captionsetup{width=\textwidth}
\includegraphics[width=\textwidth]{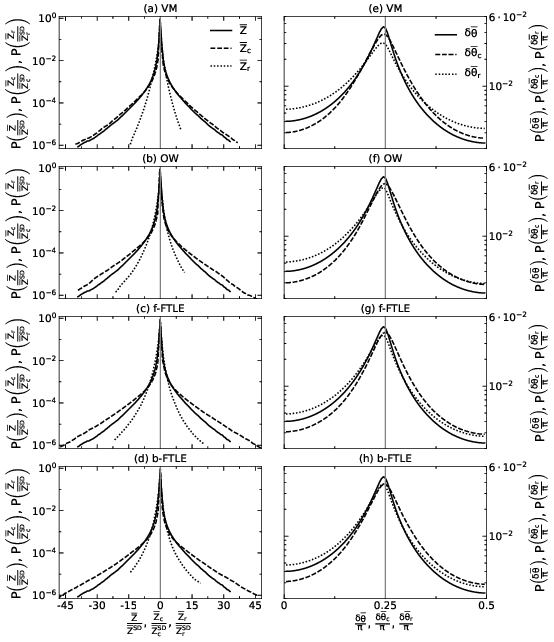}
\caption{All quantities are filtered at a wavenumber of $k=60$. 
(a)-(d): PDFs of spatial cross-scale flux distributions $\overline{Z}$, $\overline{Z}_c$ 
and $\overline{Z}_r$ with each normalised by its standard deviation 
$\overline{Z}^{SD}$, $\overline{Z}_c^{SD}$ and $\overline{Z}_r^{SD}$, respectively. 
(e)-(h): PDFs of rotation angle distributions between strain-rate and subgrid stress 
tensors $\delta \overline{\theta}$, $\delta \overline{\theta}_c$ and 
$\delta \overline{\theta}_r$.}
\label{fig:PDF_rotationangle_and_scale_to_scale_flux_vorticity_threshold_fwdFTLE_bkwdFTLE_okuboweiss_k_60_run2_3}
\end{figure}

Another observation, in the context of the present work, is that coherent structures, independent of their detection method, 
have much higher local cross-scale flux amplitudes compared to their residual counterparts. 
Using the VM criterion as an example, 
the ratio between the maximum absolute values of the 
$\omega_c$-reconstructed and the $\omega_r$-reconstructed spatial cross-scale fluxes
is $\frac{\max(|\overline{Z}_c|)}{\max(|\overline{Z}_r|)} \approx 44$. 
Similar ratios are obtained for the other coherence definitions.
However, high amplitudes are not necessarily responsible for a high net inverse flux. 
As a general observation, the probability density function (PDF) of the total cross-scale flux $\overline{Z}$,
used for reference for the comparison to the various coherence schemes shown in 
figures \ref{fig:PDF_rotationangle_and_scale_to_scale_flux_vorticity_threshold_fwdFTLE_bkwdFTLE_okuboweiss_k_60_run2_3} (a)-(d),
centralises around zero with a slightly negative skewness of $\tilde{\mu}_3 = -0.93$. 
Furthermore, the PDF has a kurtosis of $\tilde{\mu}_4 = 336$ corresponding to a very flat distribution, 
which indicates that rare high negative amplitude events contribute to the total net negative flux. 
The PDFs of the coherent parts $\overline{Z}_c$ of all coherence definitions exhibit larger tails to both positive 
and negative values compared to the total field, with kurtosis values of $404$, $430$, $503$ and $547$ 
for the VM, OW criterion, f-FTLE, and b-FTLE, respectively.
This suggests that coherent structures are responsible for the 
high spatial fluctuations of the total cross-scale flux distribution,
although these strong fluctuations do not sustain a cascade 
as already seen before with the overall low Fourier cross-scale flux contributions
in figures 
\ref{fig:comparison_decomposed_smooth_sharp_scale_energy_flux_vorticity_threshold_fwdFTLE_bkwdFTLE_okuboweiss_run2_4} 
(a)-(d).
In contrast, the PDFs of the residual contributions $\overline{Z}_r$ possess lower tails to both
positive and negative values. The negative skewness values of $-0.76$, $-1.4$, $-2.01$, and $-0.96$ 
for the VM, OW criterion, f-FTLE, and b-FTLE, respectively,
lead to a measurable net inverse cascade of the residual part.
However, due to the overall flat nature of the residual PDFs, 
the residual cascade is also driven by rare high negative amplitude events.

The angle distributions between stress and strain-rate tensors as a flux transfer efficiency measure, 
motivated by equations \eqref{eq:polar_spatial_scale_flux} and \eqref{eq:decomposed_polar_spatial_scale_flux}, 
is shown in figure \ref{fig:various_physical_quantities_run2_4} (f),
where the largest part of the  marked coherent regions display angles close to $\delta \overline{\theta} = \pi/4$. The corresponding misalignment 
of strain-rate and subfilter stress tensors results in a lower nonlinear flux efficiency.
A clearer picture of the overall cascade direction tendencies is obtained from the 
PDFs of $\delta \overline{\theta}$, $\delta \overline{\theta}_c$ and $\delta \overline{\theta}_r$ shown in 
figures \ref{fig:PDF_rotationangle_and_scale_to_scale_flux_vorticity_threshold_fwdFTLE_bkwdFTLE_okuboweiss_k_60_run2_3} (e)-(h). 
The PDF of the total field $\delta \overline{\theta}$ clearly shifts towards values smaller than $\pi/4$, 
with skewness and kurtosis values of $\tilde{\mu}_3 = 0.06$ and $\tilde{\mu}_4 = 3.79$, respectively. 
This indicates that the strain-rate and stress tensors, $\overline{\mathsfbi{S}}$ and $\overline{\boldsymbol{\tau}}$, 
tend to positively align, which lead to the overall net negative flux. 
However, the shape of all the PDFs of the coherent parts $\delta \overline{\theta}_c$ are mostly symmetric and 
centered at a value of $\pi/4$ with skewness values $0.06$, $0.02$, $0.0$, and $0.03$ for the
VM, OW criterion, f-FTLE and b-FTLE, respectively. Thus, the coherent strain-rate and stress tensors, 
$\overline{\mathsfbi{S}}_c$ and $\overline{\boldsymbol{\tau}}_{c,c}$, have the tendency to fully misalign, 
resulting in a lower cross-scale flux efficiency, despite their high local flux amplitudes. 
On the contrary, the PDFs of the residual parts $\delta \overline{\theta}_r$ 
skew even more to the right compared to the 
total field $\delta \overline{\theta}$, with skewness values of $0.14$, $0.13$, $0.12$, and $0.11$ 
for the VM, OW criterion, f-FTLE and b-FTLE, respectively. 
Additionally, for the OW criterion in 
figure \ref{fig:PDF_rotationangle_and_scale_to_scale_flux_vorticity_threshold_fwdFTLE_bkwdFTLE_okuboweiss_k_60_run2_3} (f),
for the f-FTLE in 
figure \ref{fig:PDF_rotationangle_and_scale_to_scale_flux_vorticity_threshold_fwdFTLE_bkwdFTLE_okuboweiss_k_60_run2_3} (g),
and for the b-FTLE in 
figure \ref{fig:PDF_rotationangle_and_scale_to_scale_flux_vorticity_threshold_fwdFTLE_bkwdFTLE_okuboweiss_k_60_run2_3} (h), 
the left tails of the residual parts $\delta \overline{\theta}_r$ are even higher and the right tails even 
lower compared to their corresponding coherent counterparts $\delta \overline{\theta}_c$. 
These tendencies lead to a higher stress-strain tensor alignment behavior and thus 
a higher cross-scale flux efficiency of the residual compared to the coherent part.

We conclude that although strong nonlinear cross-scale flux interactions 
are found within the quadrupolar structures of figure 
\ref{fig:coherent_residual_spatial_scale_flux_distribution_vorticity_threshold_k_60_run2_2} (a), 
these high local flux amplitudes are not responsible for an actual cascade process. 
They are rather responsible for sustaining the structures' coherence in a turbulent environment by
keeping the associated energy close to specific length scales. 

\subsection{Coherent shape preservation and residual thinning mechanism}\label{section6_2}
\begin{figure}
\centering
\captionsetup{width=\textwidth}
\includegraphics[width=\textwidth]{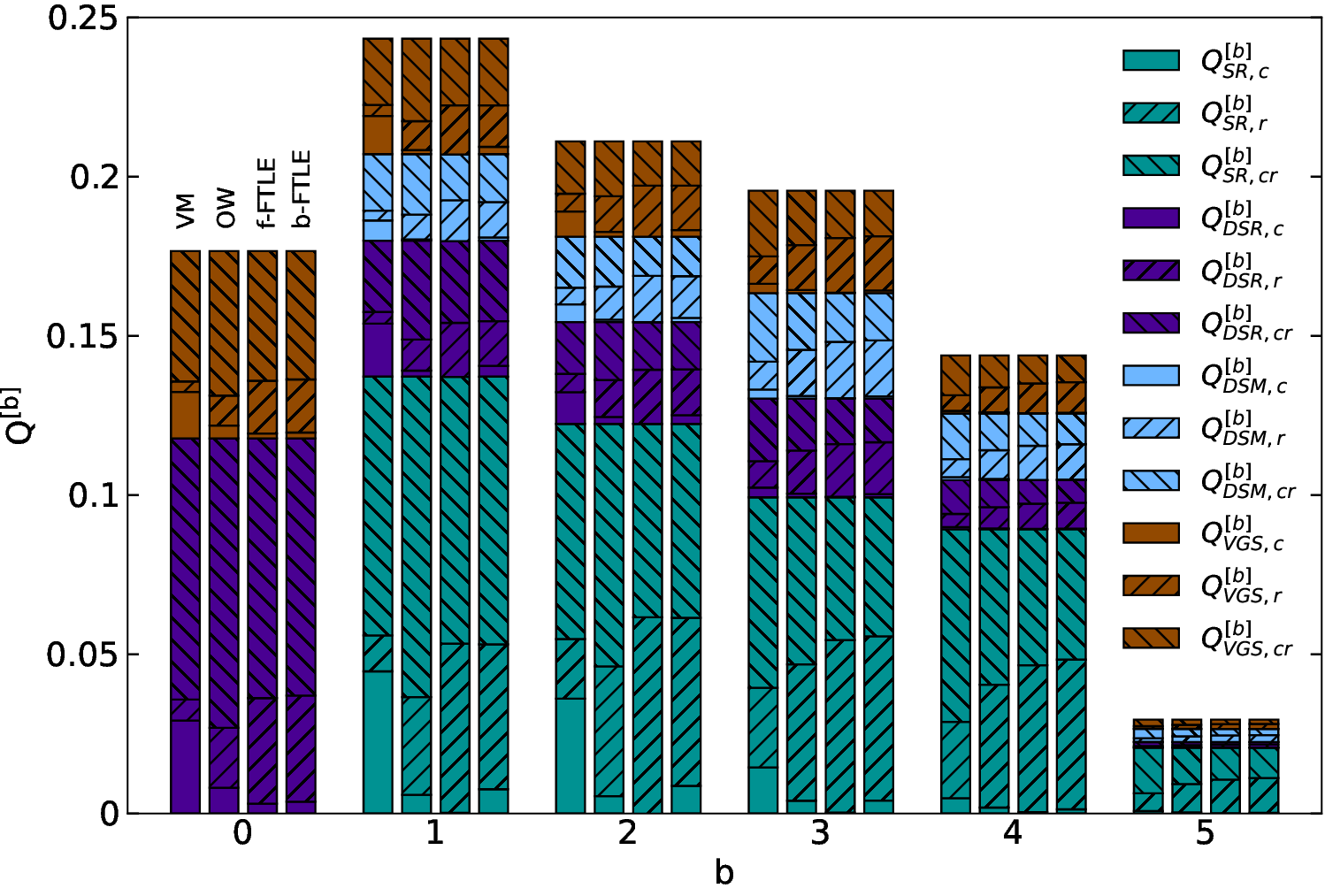}
\caption{Second-order MSG flux fractions for the length scale bands in the range of 
$b \in [0,5]$ originating from SR $Q_{SR}^{[b]}$ (green), 
DSR $Q_{DSR}^{[b]}$ (violet), 
DSM $Q_{DSM}^{[b]}$ (blue), 
and VGS $Q_{VGS}^{[b]}$ (orange) contributions, 
decoded into their corresponding coherent (subscript $c$), residual (subscript $r$), 
and mixed (subscript $cr$) parts. The varying coherence criterions 
are compared for every $b$ in the following order: 
VM, OW criterion, f-FTLE and b-FTLE.}
\label{fig:2nd_order_MSG_flux_fractions_vorticity_threshold_fwdFTLE_bkwdFTLE_okuboweiss_comparison_run2_3}
\end{figure}

A possible reason for the small coherent Fourier cross-scale flux contributions 
is the shape preservation of coherent structures in combination with
the enhanced flux efficiency of the residual part due to increased 
vortex thinning. To verify this hypothesis, the MSG expanded cross-scale energy flux in equation \eqref{eq:MSG_CSA_flux} 
and its decomposed form in equations \eqref{eq:decomposed_MSG_CSA_flux} and \eqref{eq:coherent_residual_MSG_CSA_flux}
are investigated in more detail with a filtering length of $\ell=\pi/15$, a geometric factor of $\lambda=2$
and a total number of filter bands of $n_b=5$. 
For a detailed analysis the flux fractions $Q^{[b]}$ on different subfilter-scales are divided into
\begin{align}
 Q^{[b]} = \frac{\langle Z_{SR}^{[b]} \rangle + \langle Z_{DSR}^{[b]} \rangle + \langle Z_{DSM}^{[b]} \rangle + \langle Z_{VGS}^{[b]} \rangle}{\langle Z_{*}^{MSG} \rangle} = Q_{SR}^{[b]} + Q_{DSR}^{[b]} + Q_{DSM}^{[b]} + Q_{VGS}^{[b]},
\end{align}
where $Q_{SR}^{[b]}$, $Q_{DSR}^{[b]}$, $Q_{DSM}^{[b]}$ and $Q_{VGS}^{[b]}$ are the flux fractions 
resulting from strain rotation (SR), differential strain rotation (DSR), differential strain magnification (DSM)
and vorticity gradient stretching (VGS), respectively. All the flux channels are further decomposed into 
their coherent ($c$), residual ($r$) and mixed ($cr$) contributions, e.g. the fraction of the 
SR is decomposed as $Q_{SR}^{[b]} = Q_{SR,c}^{[b]} + Q_{SR,r}^{[b]} + Q_{SR,cr}^{[b]}$. 
All these contributions are illustrated in 
figure \ref{fig:2nd_order_MSG_flux_fractions_vorticity_threshold_fwdFTLE_bkwdFTLE_okuboweiss_comparison_run2_3}. 

As already discussed by \citet{XiaoWanChenEyink2009}, the MSG flux $Z_{*}^{MSG}$ exhibits the 
locality behavior predicted by the test-field model (TFM) closure of \citet{Kraichnan1971}. 
Thus, it is not surprising in figure \ref{fig:2nd_order_MSG_flux_fractions_vorticity_threshold_fwdFTLE_bkwdFTLE_okuboweiss_comparison_run2_3}
that the majority of the 
contributions do not result from strongly local interactions $b=0$, but rather from non-local 
interactions $b \in [1,3]$, with a decreasing influence from highly non-local interactions $b \geq 4$,
with $b$ being a measure of the scale locality of nonlinear interactions.
The first-order MSG flux only consists of the SR term $Z_{SR}^{[b]}$, which has no 
contribution to the strongly local interactions ($Q_{SR}^{[0]}=0$) \citep{XiaoWanChenEyink2009,Eyink2006b}. 
Thus, second-order contributions are necessary to sufficiently capture the cross-scale flux. 
The differential strain magnification $Z_{DSM}^{[b]}$ has 
the lowest overall contribution and is also non-existent for 
strongly local interactions ($Q_{DSM}^{[0]}=0$) \citep{XiaoWanChenEyink2009,Eyink2006b}. 
Hence, its effect originating from coherent and residual parts are not further analysed with 
respect to strain-rate alignment properties. 

In figure \ref{fig:2nd_order_MSG_flux_fractions_vorticity_threshold_fwdFTLE_bkwdFTLE_okuboweiss_comparison_run2_3}
the relative contributions of the coherent, residual and mixed parts to each flux channel are nearly 
independent of the scale locality $b$ and the specific coherence definition.
The simple VM method represents a notable exception generally yielding increased coherent flux contributions at the cost
of the residual fluxes. As already mentioned above, this is the consequence of the simple gauge criterion chosen in section \ref{section5_1} that is not capable of
eliminating all differences between the detection methods.
Test simulations (not shown) indicate that the VM method is more robust than the LCS techniques to variations of the threshold
value with regard to the scaling
of the vortex number density, but exhibits a stronger and monotonic impact of the threshold on the relative importance of the different MSG flux fractions
than the LCS methods.
The gauging procedure thus leads to an effective relative threshold decrease for the VM method compared to the other more complex coherence
specifications, illustrating  the difficulty of neutralizing all differences between the coherence specifications.
The flux fractions 
mostly reflect the decomposed Fourier cross-scale flux contributions 
already presented in figure 
\ref{fig:comparison_decomposed_smooth_sharp_scale_energy_flux_vorticity_threshold_fwdFTLE_bkwdFTLE_okuboweiss_run2_4} 
above. For example, the f-FTLE criterion has small coherent cross-scale flux contributions $Z_{c,c,c}^E(k)$ 
throughout the inertial range in
figure \ref{fig:comparison_decomposed_smooth_sharp_scale_energy_flux_vorticity_threshold_fwdFTLE_bkwdFTLE_okuboweiss_run2_4} (c), 
and the coherent fractions $Q_{SR,c}^{[b]}$, $Q_{DSR,c}^{[b]}$, $Q_{DSM,c}^{[b]}$, and $Q_{VGS,c}^{[b]}$ 
of the MSG flux in figure \ref{fig:2nd_order_MSG_flux_fractions_vorticity_threshold_fwdFTLE_bkwdFTLE_okuboweiss_comparison_run2_3} 
are small for all of the flux channels as well, except for the VM method as mentioned above.
We associate the lack of substantial SR and VGS of the coherent structures with their
shape preservation characteristic.
In contrast, the residual fractions $Q_{SR,r}^{[b]}$, $Q_{DSR,r}^{[b]}$, $Q_{DSM,r}^{[b]}$, and $Q_{VGS,r}^{[b]}$ 
have a much higher contribution in all these MSG flux channels, which is also reflected by the residual cross-scale flux $Z_{r,r,r}^E(k)$. 
Additionally, the coherent cross-scale flux of the VM criterion is decreasing, 
while the residual flux is increasing for higher wavenumbers in the inertial range in 
figure \ref{fig:comparison_decomposed_smooth_sharp_scale_energy_flux_vorticity_threshold_fwdFTLE_bkwdFTLE_okuboweiss_run2_4} (a). 
This is also reflected by the MSG flux fractions in figure 
\ref{fig:2nd_order_MSG_flux_fractions_vorticity_threshold_fwdFTLE_bkwdFTLE_okuboweiss_comparison_run2_3}. 
There, the coherent fractions of all MSG flux channels decrease, while the 
residual fractions increase for lower scale locality. 
These observations are a first sign that vortex thinning plays a minor role in 
coherent structures and is more active in the residual counterpart instead. 
Lastly, similar to the decomposed cross-scale fluxes in figures 
\ref{fig:comparison_decomposed_smooth_sharp_scale_energy_flux_vorticity_threshold_fwdFTLE_bkwdFTLE_okuboweiss_run2_4} (a)-(d), 
the mixed interactions $Q_{SR,cr}^{[b]}$, $Q_{DSR,cr}^{[b]}$, $Q_{DSM,cr}^{[b]}$, and $Q_{VGS,cr}^{[b]}$ 
contribute to almost half of the fractions in all MSG flux channels in 
figure \ref{fig:2nd_order_MSG_flux_fractions_vorticity_threshold_fwdFTLE_bkwdFTLE_okuboweiss_comparison_run2_3}. 
However, this does not imply that coherent-residual interactions
lead to an enhanced thinning mechanism, since the MSG cross-scale flux of the mixed
contribution $Z_{*,cr}^{MSG}$ in equation \eqref{eq:decomposed_MSG_CSA_flux}
consists of four different heterogeneous strain-rate and stress components.
We leave the analysis of these heterogeneous interactions for future investigations.

\begin{figure}
\centering
\captionsetup{width=\textwidth}
\includegraphics[width=\textwidth]{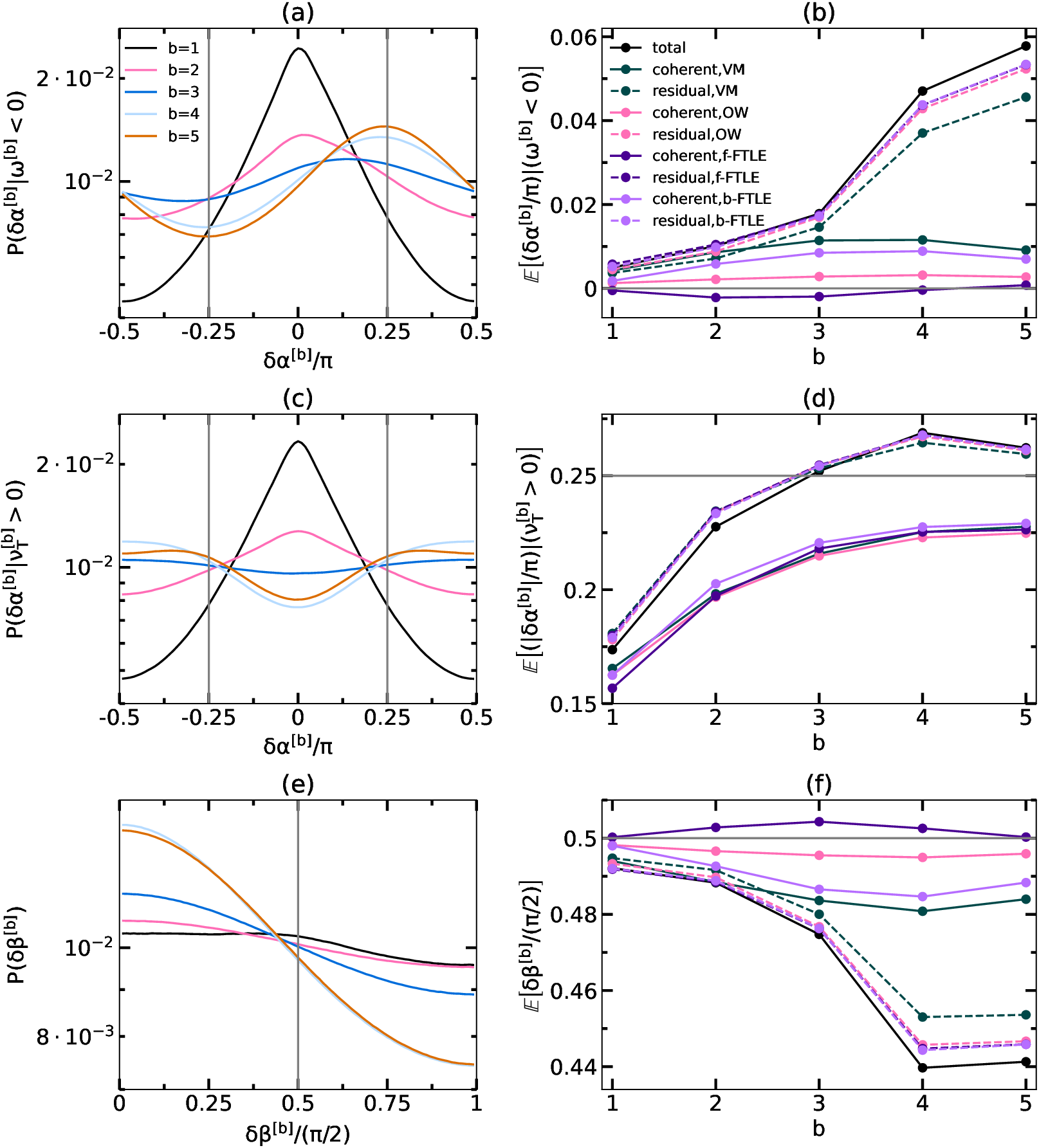}
\caption{(a): PDFs of strain-rate tensor angles $\delta \alpha^{[b]}$ conditioned 
onto the negative vorticities $\omega^{[b]}$. (b): Expected values of 
$\delta \alpha^{[b]}$, $\delta \alpha^{[b]}_c$ and $\delta \alpha^{[b]}_r$ 
conditioned onto the negative vorticities $\omega^{[b]}$, $\omega^{[b]}_c$ and 
$\omega^{[b]}_r$, respectively, for varying coherence definitions. 
(c): PDFs of strain-rate tensor angles $\delta \alpha^{[b]}$ conditioned onto the 
positive eddy-viscosities $\nu_T^{[b]}$. (d): Expected values of 
$\left|\delta \alpha^{[b]} \right|$, $\left|\delta \alpha^{[b]}_c \right|$ and 
$\left|\delta \alpha^{[b]}_r \right|$ conditioned onto the positive eddy-viscosities 
$\nu_T^{[b]}$, $\nu_{T,c}^{[b]}$ and $\nu_{T,r}^{[b]}$, respectively, 
for varying coherence definitions. (e): PDFs of angles between the large-scale 
strain-rate tensor and vorticity gradient vector $\delta \beta^{[b]}$. 
(f): Expected values of $\delta \beta^{[b]}$, $\delta \beta^{[b]}_c$ and $\delta \beta^{[b]}_r$ 
for varying coherence definitions.}
\label{fig:expected_values_conditioned_angles_2nd_order_MSG_vorticity_threshold_fwdFTLE_bkwdFTLE_okuboweiss_run2_3}
\end{figure}

According to the numerical studies of \citet{XiaoWanChenEyink2009} and as described in 
section \ref{section4_3}, the vortex thinning mechanism is geometrically quantified by 
the rotation angles $\delta \alpha^{[b]}$ and $\delta \beta^{[b]}$, which are illustrated
in figures \ref{fig:rotation_angle_overview} (b) and (c), respectively. Conditioning
the angle $\delta \alpha^{[b]}$ onto the band-pass filtered vorticities $\omega^{[b]}$
allows us to obtain statistics of the rotation angle behavior between 
the large scale strain-rate $\mathsfbi{S}^{(0)}$ and the 
band-pass filtered strain-rates $\mathsfbi{S}^{[b]}$,
and thus reveals the physical mechanism of the MSG flux contribution originated 
from the SR term $Z_{SR}^{[b]}$ in equation \eqref{eq:MSG_CSA_flux}.
A conditioning of the angle $\delta \alpha^{[b]}$ onto the band-pass filtered
eddy-viscosities $\nu_T^{[b]}$ allows the analysis of the alignment behavior
between the large scale strain-rate $\mathsfbi{S}^{(0)}$ and the band-pass
filtered stress $\boldsymbol{\tau}^{[b]} = -\nu_T^{[b]} \mathsfbi{S}^{[b]}$, which 
comes from a Newtonian stress-strain relation as already described in section \ref{section4_3}.
From this we are able to infer the physical
mechanism of the DSR term $Z_{DSR}^{[b]}$ in equation \eqref{eq:MSG_CSA_flux},
which quantifies the direction of the stress work exertion on the strain-rate field
from small to large length scales. Finally, analysing the $\delta \beta^{[b]}$ angle
reveals the influence of the VGS mechanism
described by $Z_{VGS}^{[b]}$ in equation \eqref{eq:MSG_CSA_flux}.
We extend the angle analysis by investigating the angles 
$\delta \alpha_{c/r}^{[b]}$ and $\delta \beta_{c/r}^{[b]}$
of the purely coherent and residual contributions
based on the equations \eqref{eq:decomposed_polar_coordinated_first} - \eqref{eq:decomposed_polar_coordinated_last}.
According to figures 
\ref{fig:expected_values_conditioned_angles_2nd_order_MSG_vorticity_threshold_fwdFTLE_bkwdFTLE_okuboweiss_run2_3}
(b), (d) and (f)
all coherence definitions provide similar results regarding the above mentioned angle statistics. 
Thus, the following observations and conclusions are made for coherent structures in general, 
independent of the concrete definition.

The closer the angles between the strain-rate tensors $\delta \alpha^{[b]}$ are 
to values of $\pm \pi/4$ (their sign reflecting the sign of $\omega^{[b]}$), 
the higher the  contribution of the SR term $Z_{SR}^{[b]}$ to the inverse cascade.
This rotation angle tendency is exemplarily illustrated in figure 
\ref{fig:expected_values_conditioned_angles_2nd_order_MSG_vorticity_threshold_fwdFTLE_bkwdFTLE_okuboweiss_run2_3} (a)
for the negative vorticity condition $-\omega^{[b]}$, for which the peak of the 
PDF $P(\delta \alpha^{[b]} | \omega^{[b]} < 0)$ gradually shifts towards $+\pi/4$ with 
decreasing scale locality, implied by the increasing $b$-values \citep[cf.][]{XiaoWanChenEyink2009}. 
This gradual peak shift quantifies SR and the accompanied transformation of circularly 
shaped structures into shear layers. Thus, investigating the presence of SR is one 
possibility to show the existence of the thinning process for the total field. 
Therefore, if we measure the shift of the PDF peak,
we are able to determine the SR effect separately for the purely coherent 
and residual parts. We achieve this by evaluating the conditional expectation values 
of the angles $\delta \alpha^{[b]}$, $\delta \alpha^{[b]}_c$ and $\delta \alpha^{[b]}_r$ as
\begin{align}
\mathbb{E} \left[\delta \alpha^{[b]} | (\omega^{[b]} < 0) \right], \mathbb{E} \left[\delta \alpha^{[b]}_c | (\omega^{[b]}_c < 0) \right], \mathbb{E} \left[\delta \alpha^{[b]}_r | (\omega^{[b]}_r < 0) \right].
\end{align}
Figure \ref{fig:expected_values_conditioned_angles_2nd_order_MSG_vorticity_threshold_fwdFTLE_bkwdFTLE_okuboweiss_run2_3} (b) 
shows that the conditional expectation values for the coherent structures,
independent of the concrete definition, are approximately
$\mathbb{E} \left[\delta \alpha^{[b]}_c | (\omega^{[b]}_c < 0) \right] \approx 0$.
This means that there is only a minor gradual shift of the PDF peak 
for coherent structures with decreasing scale locality (increasing $b$-values). 
This leads to the conclusion that SR and therefore thinning effects are barely 
present within structures of the coherent part. Hence, coherent structures are generally
not turned into shear layers and tend to preserve their shape.
On the contrary, the residual part exhibits an increased expected value 
$\mathbb{E} \left[\delta \alpha^{[b]}_r | (\omega^{[b]}_r < 0) \right]$
with decreasing scale locality. 
Therefore, the background residual field is prone to the development of shear layers 
and thus has a higher thinning tendency compared to the coherent structures in the system.

The DSR term $Z_{DSR}^{[b]}$ has an increasing contribution to the overall inverse cascade, 
if the angle is $\left| \delta \alpha^{[b]} \right| \gtrless \pi/4$ for
positive or negative eddy-viscosities $\pm \nu_T^{[b]}$, respectively. Therefore, 
the band-pass filtered stress tensors $\boldsymbol{\tau}^{[b]}$ should exhibit parallel or anti-parallel alignments 
to the large-scale strain-rate tensor $\mathsfbi{S}^{(0)}$ based on the eddy-viscosity's sign in order 
to maximise the $Z_{DSR}^{[b]}$ contribution. The alignment tendency reveals 
the energy transfer between scales during the vortex thinning procedure.
Thus, negative work exertion of the small-scale stress $\boldsymbol{\tau}^{[b]}$ on the large-scale
strain-rate field $\mathsfbi{S}^{(0)}$ is understood as a general energy transfer 
from small to large length scales enabling an overall inverse energy cascade,
whereas positive work exertion allows for a direct energy cascade.
The PDF of the angles exemplarily conditioned onto the positive 
eddy-viscosity $P(\delta \alpha^{[b]}|\nu_T^{[b]} > 0)$ 
for the total field is presented in figure 
\ref{fig:expected_values_conditioned_angles_2nd_order_MSG_vorticity_threshold_fwdFTLE_bkwdFTLE_okuboweiss_run2_3} (c)
and shows that the angle becomes 
$\left| \delta \alpha^{[b]} \right| > \pi/4$ for $b \geq 3$ \citep[cf.][]{XiaoWanChenEyink2009}. 
This reveals the negative work exertion of the small-scale stress $\boldsymbol{\tau}^{[b]}$ 
on the large-scale strain-rate tensor $\mathsfbi{S}^{(0)}$, as depicted in figure 
\ref{fig:rotation_angle_overview} (b), and leads to the
conclusion that the vortex thinning mechanism facilitates the energy
transfer from small to large length scales for the entire field. The conditional expectation values
of the angles $\delta \alpha^{[b]}$, $\delta \alpha^{[b]}_c$ and $\delta \alpha^{[b]}_r$
\begin{align}
 \mathbb{E} \left[\left| \delta \alpha^{[b]} \right| | (\nu_T^{[b]} > 0) \right], \mathbb{E} \left[\left| \delta \alpha^{[b]}_c \right| | (\nu_{T,c}^{[b]} > 0) \right], \mathbb{E} \left[\left| \delta \alpha^{[b]}_r \right| | (\nu_{T,r}^{[b]} > 0) \right],
\end{align}
are determined to measure the shift of the maxima towards $\left| \delta \alpha^{[b]} \right| > \pi/4$ 
for increasing $b$-values 
in figure \ref{fig:expected_values_conditioned_angles_2nd_order_MSG_vorticity_threshold_fwdFTLE_bkwdFTLE_okuboweiss_run2_3} (d) 
and are also used to determine the alignment tendencies of coherent and residual parts. 
The absolute values $\left| \delta \alpha^{[b]} \right|$ 
instead of the original angle $\delta \alpha^{[b]}$ are considered, because the PDFs in 
figure \ref{fig:expected_values_conditioned_angles_2nd_order_MSG_vorticity_threshold_fwdFTLE_bkwdFTLE_okuboweiss_run2_3} (c) 
are symmetric. As a result, 
figure \ref{fig:expected_values_conditioned_angles_2nd_order_MSG_vorticity_threshold_fwdFTLE_bkwdFTLE_okuboweiss_run2_3} (d) 
shows that the negative work exertion is present within the residual parts for $b \geq 3$ 
as the angles become $\left| \delta \alpha^{[b]}_r \right| > \pi/4$.
Although $\mathbb{E} \left[\left| \delta \alpha^{[b]}_c \right| | (\nu_{T,c}^{[b]} > 0) \right]$ 
increases for larger $b$, the angles for the coherent structures remain at 
$\left| \delta \alpha^{[b]}_c \right| < \pi/4$
independent of the scale locality.
This means that the coherent part even has the tendency
that the coherent band-pass filtered stress tensors $\boldsymbol{\tau}_c^{[b]}$ exert 
positive instead of negative work on the coherent large-scale strain-rate tensor 
$\mathsfbi{S}_c^{(0)}$. Thus, energy within coherent structures is
not transferred from the small-scale stress to the large-scale 
strain-rate due to the weak contributions of the thinning mechanism.

Lastly, the VGS term $Z_{VGS}^{[b]}$ is dependent 
on the angles $\delta \beta^{[b]}$ between the contractile direction 
of the large-scale strain-rate $\mathsfbi{S}^{(0)}$ and the 
vorticity gradients $\nabla \omega^{[b]}$. These angles approach zero $\delta \beta^{[b]} \rightarrow 0$ 
for decreasing scale locality (increasing $b$), as presented in figure 
\ref{fig:expected_values_conditioned_angles_2nd_order_MSG_vorticity_threshold_fwdFTLE_bkwdFTLE_okuboweiss_run2_3} (e) 
\cite[cf.][]{XiaoWanChenEyink2009},
which shows the contribution $Z_{VGS}^{[b]}$ for the total field.
This quantifies the presence of the vorticity isoline deformation 
along the stretching direction of the large-scale strain-rate field, 
as depicted in figure \ref{fig:rotation_angle_overview} (b) for the total field. 
The alignment angles $\delta \beta_{c/r}^{[b]}$ between 
the purely coherent/residual vorticity gradients $\nabla \omega_{c/r}^{[b]}$
and the purely coherent/residual contractile direction
of the strain-rate tensor $\mathsfbi{S}_{c/r}^{(0)}$
are used to measure the effect of VGS in the coherent and
residual parts of the flow respectively.
The following expected values
\begin{align}
 \mathbb{E} \left[\delta \beta^{[b]} \right], \mathbb{E} \left[\delta \beta^{[b]}_c \right], \mathbb{E} \left[\delta \beta^{[b]}_r \right],
\end{align}
for the total, coherent and residual fields are illustrated in figure 
\ref{fig:expected_values_conditioned_angles_2nd_order_MSG_vorticity_threshold_fwdFTLE_bkwdFTLE_okuboweiss_run2_3} (f). 
The expected value for the residual angle $\mathbb{E} \left[\delta \beta^{[b]}_r \right]$ 
decreases for large scale separations (increasing $b$), indicating the presence of 
the VGS in the residual part and thus is another indicator of the thinning mechanism.
For the coherent part $\mathbb{E} \left[\delta \beta^{[b]}_c \right]$
is close to $\pi/4$ for all values of $b$. This implies the physical picture that the 
coherent large-scale strain-rate tensor $\mathsfbi{S}_{c}^{(0)}$ is not distorting the
vorticity isolines, which ultimately leads to a shape preservation of coherent structures.

In conclusion, the small cascade efficiency of 
coherent structures is caused by their shape preserving nature determined 
by the depletion of $Z_{SR,c}^{[b]}$ and $Z_{VGS,c}^{[b]}$ terms independent of the 
scale separations $b$ in combination with the positive work exertion of the small-scale 
stresses on the large scale strain-rate, as captured by the $Z_{DSR,c}^{[b]}$ 
term. In contrast, the higher flux efficiency in the residual parts of the flow is caused 
by the enhanced thinning mechanism quantified by the 
$Z_{SR,r}^{[b]}$ and $Z_{VGS,r}^{[b]}$ contributions
and the negative work exertion captured in the $Z_{DSR,r}^{[b]}$ term.

\section{Conclusions}\label{section7}
The present investigation deals with the nonlinear dynamics of 
coherent and residual structures in
pseudospectral DNS of two-dimensional Navier-Stokes turbulence forced
at small scales. This involves the application of three commonly applied coherence detection
schemes based on the OW criterion, on the VM, and on
LCSs to identify and to isolate the respective flow components. Among these threshold-based
detection methods, the VM technique and the 
LCS approach are gauged by using statistical properties of
detected structures 
to improve comparability of the detection results. Using this setup, 
$(i)$ the performance of the employed detection
methods is discussed in relation to each other,
$(ii)$ the coherent and residual contributions to the cross-scale energy flux 
of the inverse turbulent cascade 
are analyzed in spectral Fourier as well as in configuration space to study their 
role for the characteristics and for the
dynamics of the respective parts of the flow.

We find that, $(i)$, under application of the chosen gauge criterion and in 
comparison with the VM scheme the OW method exhibits a bias towards
largest-scale and most energetic structures. In contrast, the LCS scheme shows 
an increased susceptibility for small-scale coherence
as compared to the VM method. Both tendencies can largely
be neutralized by adjusting the free parameters of the VM and the
LCS methods to yield the same scale-dependency of the vortex number
density as the OW specification.

With respect to the role of detected coherent structures for turbulence dynamics, $(ii)$, we
find that they are responsible for a pollution of the
phenomenologically expected spectral scaling of the kinetic energy
spectrum $E(k)\sim k^{-5/3}$ at largest spatial scales. This finding
is supported by the largely unaffected $k^{-5/3}$-scaling observed in
the energy spectrum of the residual (incoherent) fraction of the
turbulent fluctuations. The observation suggests the possible use of
coherence detection and decomposition in DNS
of homogeneous turbulence for the reduction of the large-scale
condensation of inversely cascading quantities for physical systems
that feature inverse cascades, e.g. two-dimensional Navier-Stokes
turbulence or magnetohydrodynamic turbulence.

The application of a spatial scale-filter approach for the analysis of
the nonlinear dynamics of coherent and residual parts of the turbulent
flow indicates a high nonlinear activity within
coherent structures.
The finding shows that coherent structures in two-dimensional Navier-Stokes
turbulence are in general dynamically sustained while the spatial
structure of the dynamics yields a shape-preserving depletion of the
nonlinear cross-scale flux with regard to the entire structure.
This is in agreement with the observed coherent 
Fourier cross-scale energy fluxes and the low flux efficiency
due to the high misalignment tendencies of
coherent strain-rate and subgrid stress tensors.
The shape preservation of coherent structures in this case is verified
by employing the 
MSG expansion of the coherent spatial cross-scale energy fluxes that exhibits
a clear depletion of the deformation processes that are scale-flux generating.
These findings suggest to employ the depletion of the MSG contributions of SR and
VGS as markers for structural coherence in two-dimensional turbulent flows.

The inverse cascade is instead driven by a combination of
$(i)$ interactions entirely among residual fluctuations and of $(ii)$ nonlinear
interactions between coherent structures and residual fluctuations.
The former contribution is strongest at small scales while
the latter dominates at large scales. This suggests that two different
physical processes are responsible for the respective energy fluxes. 
For the first contribution the dominant physical process
has recently been introduced as vortex thinning.
This is in line with the enhanced alignment properties
of the residual strain-rate and subgrid stress tensors, yielding
a high flux efficiency.
The second contribution is the dominant flux contribution
and stays on a relatively high and roughly constant level 
thoughout the inverse flux region.
We abstain from ascribing this flux contribution to more 
specific physical dynamics as multiple factors may be determining 
its characteristics due to the heterogeneous character of the 
interacting strain-rate and and stress tensor fields.
Further work is presently being pursued along these directions.

\section*{Acknowledgements}
The authors thank B. Beck, R. M\"ausle, J. Reiss,
and J.-M. Teissier for fruitful discussions.
This work was supported by the German Research Foundation (DFG) 
within the Research Training Group GRK 2433. Computing
resources from the Max Planck Computing and Data Facility (MPCDF) 
are also acknowledged.

\noindent{\bf Declaration of Interests}. The authors report no conflict of interest.

\appendix
\section{Technical details}\label{appendix1}

\subsection{Vorticity decomposition}\label{appendix1_1}
After thresholding according to one of the investigated coherence
criteria in the fashion of equations \eqref{eq:decomposition_coherent}
and \eqref{eq:decomposition_residual}, the nonzero scalar values of
the vorticity field are clustered. All adjacent neighboring nonzero
pixels (maximum of four neighbors for each pixel) are grouped into
separately countable and connected clusters. Then, clusters whose
number of pixels are below the forcing area of $A_f = \pi (\ell_f/2)^2$,
with $\ell_f = 2 \pi/k_f$ the forcing length scale, are excluded and
discarded from the coherent field. After that, a $5 \times 5$ smooth
Gaussian filter is applied to avoid regularity issues caused by
sharp boundaries. This leads to the coherent vorticity field
$\omega_c^{smooth}$. The residual field is obtained by 
subtraction,

\begin{align}
 \omega_r = \omega - \omega_c^{smooth}.
\end{align}

\subsection{Efficient FTLE calculation}\label{appendix1_2}
From a technical viewpoint, integrating larger amounts of Lagrangian
tracers to sufficiently resolve a turbulent flow in space and time 
requires substantial computational resources. 
Thus, we utilize an efficient
numerical technique, as suggested by \citet{FinnApte2013}, to simultaneously 
calculate the flowmap $\boldsymbol{F}_{t_0}^{t}$ forward and backward
in time by employing a Lagrangian and a semi-Lagrangian scheme, respectively.
The forward-in-time/Lagrangian scheme advects passive
tracers for an integration time $T$ according to the underlying
velocity field, which generate the forward-in-time flowmap, $\boldsymbol{F}_{t_0}^{t_0+T}$. 
The backward-in-time/semi-Lagrangian scheme, introduced by \citet{Leung2011}, 
is based on the solution of level-set equations
and constructs the backward-in-time flowmap, $\boldsymbol{F}_{t_0}^{t_0-T}$, 
by tracking coordinates on an Eulerian grid.

For an increased temporal resolution of the FTLE fields, a flowmap
composition method, proposed by \citet{BruntonRowley2010}, is applied as
\begin{align}
\boldsymbol{F}_{t_0}^{t_0+T} = \mathcal{I} \boldsymbol{F}_{t_0+(N-1)h}^{t_0+Nh} \circ \ldots \circ \mathcal{I} \boldsymbol{F}_{t_0+h}^{t_0+2h} \circ \boldsymbol{F}_{t_0}^{t_0+h},
\end{align}
where $\circ$ is the composition operator, $N$ the number of substeps,
$h=100$ the flowmap substep and $\mathcal{I}$ the interpolation
operator. A second-order Runge-Kutta integration scheme is applied for
particle integration and a cubic interpolation scheme, 
suggested by \citet{StaniforthCote1991}, is used for the
interpolation operator and the mapping of particles between grid
points.

\subsection{MSG flux derivation}\label{appendix1_3}
The multi-gradient nature of the approach arises from a Taylor expansion 
\begin{align}
\delta \boldsymbol{u}^{(b,m)}(\boldsymbol{r};\boldsymbol{x}) = \sum_{p=1}^{m} \frac{1}{p!} (\boldsymbol{r} \bcdot \nabla)^p \boldsymbol{u}^{(b)}(\boldsymbol{x})
\end{align}
of the filtered velocity increments
$\delta \boldsymbol{u}^{(b)}(\boldsymbol{r};\boldsymbol{x})
= \boldsymbol{u}^{(b)}(\boldsymbol{x}+\boldsymbol{r}) - 
\boldsymbol{u}^{(b)}(\boldsymbol{x})$ with separation
vector $\boldsymbol{r}$,
which is subject to the filtering operation given by equation 
\eqref{eq:multiscale_filtered_velocity}. 
The filtered subgrid stress tensor $\boldsymbol{\tau}^{(b)}$
can then be entirely expressed by the Taylor expanded velocity
increments instead of the original increments $\delta
\boldsymbol{u}^{(b)}(\boldsymbol{r})$. This yields \citep[cf.][]{Eyink2006}
\begin{align}
\boldsymbol{\tau}^{(b,m)} &= \int G_l(\boldsymbol{r}) \delta \boldsymbol{u}^{(b,m)}(\boldsymbol{r}) \delta \boldsymbol{u}^{(b,m)}(\boldsymbol{r}) \diffunit^2 \boldsymbol{r} \nonumber \\ 
&- \int G_l(\boldsymbol{r}) \delta \boldsymbol{u}^{(b,m)}(\boldsymbol{r}) \diffunit^2 \boldsymbol{r} \int G_l(\boldsymbol{r}) \delta \boldsymbol{u}^{(b,m)}(\boldsymbol{r}) \diffunit^2 \boldsymbol{r}, 
\end{align}
where $\boldsymbol{\tau}^{(b,m)}$ is a multi-scale and multi-gradient expression for the stress.

It can be shown that the MSG expanded stress
converges as $\lim_{m \rightarrow \infty} \boldsymbol{\tau}^{(b,m)} =
\boldsymbol{\tau}^{(b)}$ in the $L^1$-norm. Nevertheless, for
{increasing scale indices $b$}, 
which corresponds to adding finer-scale
structures, a growing amount of space gradients of higher-order, $m$,
is required to approximate $\boldsymbol{\tau}^{(b,m)} \approx
\boldsymbol{\tau}^{(b)}$. Therefore, a \textit{coherent-subregions
approximation} (CSA) approach is suggested by \citet{Eyink2006}
enabling the approximation of the MSG stress by low-order gradients
$m$. As a result, the CSA corrected MSG stress is obtained, which is
more accurately approximated as $\boldsymbol{\tau}^{(b,m)}_{*} \approx
\boldsymbol{\tau}^{(b,m)}$ with fewer gradients $m$. According to
\citet{Eyink2006b} and \citet{XiaoWanChenEyink2009}, the best approximation with
regard to the original cross-scale flux term is reached for expansions up to
the second-order in gradients $m=2$. For the sake of brevity we define
$\boldsymbol{\tau}_{*}^{MSG}=\boldsymbol{\tau}_{*}^{(b,2)}$ and $Z_{*}^{MSG} =
Z_{*}^{(b,2)}$, which are the MSG-CSA stress and cross-scale flux,
respectively, expanded to second order, giving
\begin{align}
 Z_{*}^{MSG} &= - \mathsfbi{S}^{(0)} : \boldsymbol{\tau}_{*}^{MSG} \nonumber \\
 &= \sum_{b=0}^{n_b} \Bigg[ \underbrace{-\frac{\overline{C}_2^{[b]}}{2} \ell_b^2 \omega^{[b]} (\mathsfbi{S}^{(0)} : \tilde{\mathsfbi{S}}^{[b]})}_{Z_{SR}^{[b]}} \underbrace{ + \frac{\overline{C}_4^{[b]}}{8} \ell_b^4 (\overbrace{\nabla \omega^{[b]} \bcdot \nabla \alpha^{[b]}}^{\nu_T^{[b]}}) (\mathsfbi{S}^{(0)} : \mathsfbi{S}^{[b]})}_{Z_{DSR}^{[b]}} \nonumber \\
 &\underbrace{-\frac{\overline{C}_4^{[b]}}{16} \ell_b^4 (\overbrace{\nabla \omega^{[b]} \bcdot \nabla \ln \sigma^{[b]}}^{\gamma_T^{[b]}}) (\mathsfbi{S}^{(0)} : \tilde{\mathsfbi{S}}^{[b]})}_{Z_{DSM}^{[b]}} \underbrace{+ \frac{\overline{C}_4^{[b]}}{32} \ell_b^4 (\nabla \omega^{[b]})^{T} \mathsfbi{S}^{(0)} (\nabla \omega^{[b]})}_{Z_{VGS}^{[b]}} \Bigg] \nonumber \\
 & \underbrace{- (\nabla \psi_{*}^{(n_b)})^{T} \mathsfbi{S}^{(0)} (\nabla \psi_{*}^{(n_b)})}_{Z_{FSF}^{(n_b)}}\,, 
\end{align}
with the fluctuation stream function $\psi_{*}^{(n_b)}$
\citep[cf.][]{Eyink2006b} and the coefficients $\overline{C}_p^{[b]}$
\citep[cf. appendix C in][]{Eyink2006}.
The flux contribution from the fluctuation stream function (FSF)
$Z_{FSF}^{(n_b)}$ is similarly interpreted as a vorticity gradient
stretching but considered much smaller in magnitude due to
cancellations from spatial averaging. 
Additionally,
it possesses a positive mean as shown by \citet{XiaoWanChenEyink2009},
thus a further analysis with regard to the inverse cascade mechanism
is neglected for this term.

\section{Sensitivity of energy injection rate, KLB theory and finite-size effects}\label{appendix2}
The present coherent structure analysis is based on DNS of the
two-dimensional Navier-Stokes equations. In this section, we briefly discuss the
choice of system parameters, with regard to the dynamics of structure
formation and turbulence statistics.

\begin{figure}
\centering
\captionsetup{width=\textwidth}
\includegraphics[width=\textwidth]{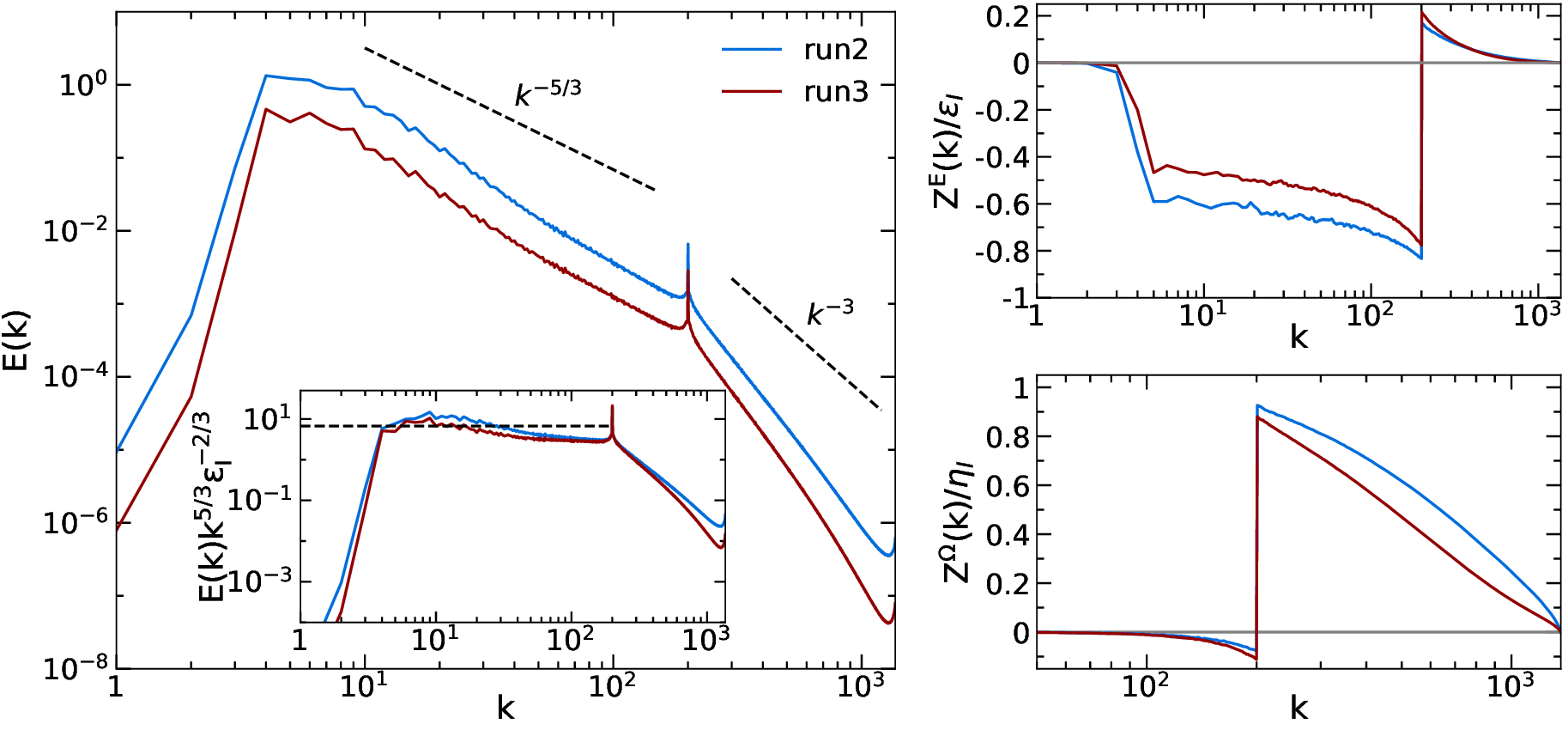}
\caption{Kinetic energy spectra $E(k)$ of run2 and 3 with varying 
kinetic energy injection rates (left) with inset showing the 
compensated spectra $E(k) k^{5/3} \epsilon_I^{-2/3}$, where the black 
dashed line indicates a value of $C_E = 6.69$ predicted by the TFM 
closure of \citet{Kraichnan1971}. Normalised cross-scale kinetic energy 
$Z^E(k)/\epsilon_I$ (top right) and enstrophy fluxes $Z^{\Omega}(k)/\eta_I$ 
(bottom right) from run2 and 3, respectively.}
\label{fig:total_kinetic_energy_spectrum_energy_enstrophy_flux_run2_run3}
\end{figure}

In contrast to other studies
\citep[see][]{Borue1994,BabianoBasdevantLegrasSadourny1987,MaltrudVallis1993,
DanilovGurarie2001,Vallgren2011,BurgessScott2018},
the present work does not employ a hyperviscous dissipation term to
avoid the accompanying unphysical distortion of the spatial structure of the flow.
Although this shortens the scaling-range of the enstrophy cascade, the
resulting spectral extent still suffices for our purposes, as shown by
the spectra and enstrophy fluxes in figures
\ref{fig:time_evolution_vorticity_spectra_fluxes_run2} (c) and (d),
and figure
\ref{fig:total_kinetic_energy_spectrum_energy_enstrophy_flux_run2_run3}.
We have found no further implications for other diagnostics relevant
in the context of the  present investigation.

The formation of structures is dependent on the energy injection rate
$\epsilon_I$ and therefore we vary its rate for fixed viscosity $\nu$
and large-scale friction values ($k_{0,\omega}$, $\sigma_{\omega}^2$,
$\alpha_{\omega}$), which also affects the {ratio} of energy
dissipated at largest scales with rate $\epsilon_{\alpha}$ and at
viscous scales with rate $\epsilon_{\nu}$. Due to the unavoidably
limited spectral bandwidth of the numerical simulations, it is not
possible to fulfill both requirements of the KLB picture at the same
time, namely a ratio of $\epsilon_{\alpha}/\epsilon_I = 1$, such that
all the injected energy is dissipated at large-scales, in combination
with an exact power-law $\sim k^{-5/3}$ for the energy
scaling-range. Even DNS at significantly higher numerical resolution
\citep[see e.g.][]{BoffettaMusacchio2010} exhibit smaller deviations from
the asymptotic scaling exponent.  We observe, that simultaneously only
one of the two characteristics is approximately realisable with
sufficient accuracy.  Therefore, three simulation runs are performed,
as listed in table
\ref{tab:simulation_parameters_inverse_kinetic_energy_cascade_hydro_runs}.
Next to numerical resolution and the large-scale damping required for quasi-stationarity of the flow,
the large-scale driving, i.e. the energy injection rate $\epsilon_I$, exerts an important influence on the system.

\begin{table}
\captionsetup{width=\textwidth}
\begin{center}
\begin{tabular}{lccccc}
Run & $\epsilon_I$ & $\epsilon_{\alpha}/\epsilon_I$ & $\epsilon_{\nu}/\epsilon_I$ &  $Re [10^{5}]$ & $T_{eddy}$ \\[3pt]
run1 & $13.05$ & $0.67$ & $0.33$ & $7.87$ & $0.1$ \\
run2 & $3.5$ & $0.6$ & $0.4$ & $4.36$ & $0.2$ \\
run3 & $0.88$ & $0.47$ & $0.53$ & $2.34$ & $0.4$ \\
\end{tabular}
\caption{Simulation parameters of the two-dimensional hydrodynamic 
turbulence system with a resolution of $4096^2$, forcing wavenumber 
$k_f = 200$, large-scale friction factor $\alpha_{\omega}=500$, 
center of Gaussian damping profile $k_{0,\omega} = 0.1$, variance 
of the Gaussian $\sigma_{\omega}^2 = 1$, viscosity 
$\nu=5 \bcdot 10^{-5}$ and timestep $dt=5 \bcdot 10^{-6}$.}
\label{tab:simulation_parameters_inverse_kinetic_energy_cascade_hydro_runs}
\end{center}
\end{table}

According to figure
\ref{fig:time_evolution_vorticity_spectra_fluxes_run2} (c), run1 with
the highest energy injection rate exhibits the strongest deviation
from the $k^{-5/3}$ scaling, but the best developed 
normalised cross-scale energy $Z^E(k)/\epsilon_I$ and
enstrophy fluxes $Z^{\Omega}(k)/\eta_I$ (figure
\ref{fig:time_evolution_vorticity_spectra_fluxes_run2} (d))
close to values of $-1$ and
$1$ in the energy and enstrophy inertial ranges, respectively. Run3
with the lowest energy injection rate in figure
\ref{fig:total_kinetic_energy_spectrum_energy_enstrophy_flux_run2_run3}
is the closest to fulfill the power-law but displays weaker nonlinear
fluxes. Although claims exist that relate the large-scale steepening
of the energy spectrum to the large-scale damping leading to the
formation of large-scale coherent vortices at largest scales
\citep[see][]{Borue1994,DanilovGurarie2001}, other studies show that vortex
formation already occurs on smaller scales and is not entirely an
artifact due to hypofriction effects
\citep[see][]{BabianoBasdevantLegrasSadourny1987,Vallgren2011,BurgessScott2017}. 
Therefore, it appears to be plausible that the formation of 
coherent structures in configuration space is an inherent 
property of two-dimensional turbulence not captured by the 
wavenumber-based KLB phenomenology. 
This structure formation property has the 
tendency to pollute the scaling of the energy spectrum, 
which is further discussed in
section \ref{section5}.

\begin{figure}
\centering
\captionsetup{width=\textwidth}
\includegraphics[width=\textwidth]{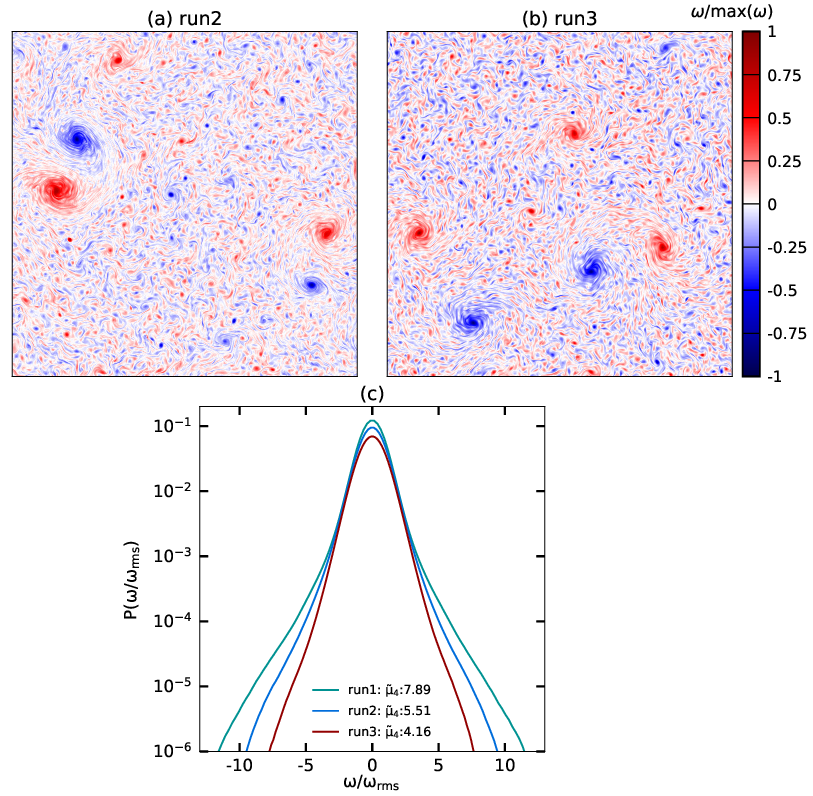}
\caption{(a), (b): Vorticities of run2 and 3 from a $1024^2$ region, 
which is $6.25\%$ of the total physical domain. The colorbar is normalised 
to $\omega/\max{|\omega|}$ for a better visual comparison of structures 
from varying injection rates. (c): The kurtosis of the vorticity PDFs 
from runs1-3 are denoted as $\tilde{\mu}_4$.}
\label{fig:comparison_vorticity_and_averaged_pdf_run2_run3_2}
\end{figure}

With decreasing energy injection rate the vorticity field has less
distinct coherent features as shown in figure
\ref{fig:time_evolution_vorticity_spectra_fluxes_run2} (b) and figures
\ref{fig:comparison_vorticity_and_averaged_pdf_run2_run3_2} (a) and (b),
where single vortex structures become less intense and gradually
dissolve into the residual background, similarly observed by
\citet{BurgessScott2017}. At the same time the probability density
function (PDF) of the vorticity becomes flatter with increasing
injection rates according to figure
\ref{fig:comparison_vorticity_and_averaged_pdf_run2_run3_2} (c), where
the increasing values at the tails of the PDFs are associated to the
large vorticity values in the visible vortex cores. For the present
analysis run1 is used, due to the most visible presence of coherent
structures and the stronger
spectral cross-scale flux in the
inverse cascade regime compared to the other DNS. However, the
remaining DNS, run2 and run3, also
lead to qualitatively similar results. This suggests 
that the similarity scaling is not the determining factor for
the inverse cascade dynamics of coherent structures.

The large-scale damping certainly has a strong and intended effect on the large-scale energetics that naturally extends over a
limited spectral range into the smaller-scale dynamics. The high level of isotropy of the damping mechanism, however, is an effort to avoid
additional severe nuisances such as violent and random artificial straining or other unwanted anisotropic processes. Although comparison with
similar works without applied large-scale damping suggests that the damping does not lead to even more severe artefacts with regard to coherence
dynamics as already inflicted by the finite-size of the system. Such unwanted side-effects in particular with regard to the higher-order MSG results
can of course not be fully ruled out.
Furthermore, the impact of numerical resolution, as observed in test simulations with a monotonically decreasing number of collocation points down
to $256^2$ reveals that
the three-regime signature of the vortex number density already becomes less discernible at a resolution of $2048^2$.
The qualitative results obtained via the MSG expansion, in contrast, remain unchanged and observable for all tested resolutions.

\bibliographystyle{jfm}
\bibliography{references}

\begin{thebibliography}{57}
\expandafter\ifx\csname natexlab\endcsname\relax\def\natexlab#1{#1}\fi
\def\au#1{#1} \def\ed#1{#1} \def\yr#1{#1}\def\at#1{#1}\def\jt#1{\textit{#1}}
  \def\bt#1{#1}\def\bvol#1{\textbf{#1}} \def\vol#1{#1} \def\pg#1{#1}
  \def\publ#1{#1}\def\arxiv#1{#1}\def\org#1{#1}\def\st#1{\textit{#1}}

\bibitem[Babiano {\em et~al.\/}(1987)Babiano, Basdevant, Legras \&
  Sadourny]{BabianoBasdevantLegrasSadourny1987}
{\sc \au{Babiano, A.}, \au{Basdevant, C.}, \au{Legras, B.} \& \au{Sadourny,
  R.}} \yr{1987}  \at{Vorticity and passive-scalar dynamics in two-dimensional
  turbulence}.  \jt{Journal of Fluid Mechanics}  \bvol{183},  \pg{379--397}.

\bibitem[Batchelor(1969)]{Batchelor1969}
{\sc \au{Batchelor, G.~K.}} \yr{1969}  \at{Computation of the energy spectrum
  in homogeneous two-dimensional turbulence}.  \jt{The Physics of Fluids}
  \bvol{12},  \pg{233--239}.

\bibitem[Benzi {\em et~al.\/}(1986)Benzi, Paladin, Patarnello, Santangelo \&
  Vulpiani]{BenziPaladinPatarnelloSantangeloVulpiani1986}
{\sc \au{Benzi, R.}, \au{Paladin, G.}, \au{Patarnello, S.}, \au{Santangelo, P.}
  \& \au{Vulpiani, A.}} \yr{1986}  \at{Intermittency and coherent structures in
  two-dimensional turbulence}.  \jt{Journal of Physics A: Mathematical and
  General}  \bvol{19},  \pg{3771--3784}.

\bibitem[Benzi {\em et~al.\/}(1988)Benzi, Patarnello \&
  Santangelo]{BenziPatarnelloSantangelo1988}
{\sc \au{Benzi, R.}, \au{Patarnello, S.} \& \au{Santangelo, P.}} \yr{1988}
  \at{Self-similar coherent structures in two-dimensional decaying turbulence}.
   \jt{Journal of Physics A: Mathematical and General}  \bvol{21},
  \pg{1221--1237}.

\bibitem[Boffetta \& Musacchio(2010)]{BoffettaMusacchio2010}
{\sc \au{Boffetta, G.} \& \au{Musacchio, S.}} \yr{2010}  \at{Evidence for the
  double cascade scenario in two-dimensional turbulence}.  \jt{Physical Review
  E}  \bvol{82},  \pg{016307}.

\bibitem[Borue(1994)]{Borue1994}
{\sc \au{Borue, V.}} \yr{1994}  \at{Inverse energy cascade in stationary
  two-dimensional homogeneous turbulence}.  \jt{Physical Review Letters}
  \bvol{72}~(10),  \pg{1475--1478}.

\bibitem[Brunton \& Rowley(2010)]{BruntonRowley2010}
{\sc \au{Brunton, S.~L.} \& \au{Rowley, C.~W.}} \yr{2010}  \at{Fast computation
  of finite-time lyapunov exponent fields for unsteady flows}.  \jt{Chaos}
  \bvol{20},  \pg{017503}.

\bibitem[Burgess \& Scott(2017)]{BurgessScott2017}
{\sc \au{Burgess, B.~H.} \& \au{Scott, R.~K.}} \yr{2017}  \at{Scaling theory
  for vortices in the two-dimensional inverse energy cascade}.  \jt{Journal of
  Fluid Mechanics}  \bvol{811},  \pg{742--756}.

\bibitem[Burgess \& Scott(2018)]{BurgessScott2018}
{\sc \au{Burgess, B.~H.} \& \au{Scott, R.~K.}} \yr{2018}  \at{Robustness of
  vortex populations in the two-dimensional inverse energy cascade}.
  \jt{Journal of Fluid Mechanics}  \bvol{850},  \pg{844--874}.

\bibitem[Canivete~Cuissa \& Steiner(2020)]{CaniveteCuissaSteiner2020}
{\sc \au{Canivete~Cuissa, J.~R.} \& \au{Steiner, O.}} \yr{2020}  \at{Vortices
  evolution in the solar atmosphere: A dynamical equation for the swirling
  strength}.  \jt{Astronomy and Astrophysics}  \bvol{639},  \pg{A118}.

\bibitem[Canuto {\em et~al.\/}(1988)Canuto, Hussaini, Quarteroni \&
  Zang]{CanutoHussainiQuarteroniZang_SpectralMethodsInFluidDynamics_1988}
{\sc \au{Canuto, C.}, \au{Hussaini, M.~Y.}, \au{Quarteroni, A.} \& \au{Zang,
  T.~A.}} \yr{1988} {\em Spectral methods in fluid dynamics\/}.
  \publ{Springer-Verlag Berlin Heidelberg}.

\bibitem[Chakraborty {\em et~al.\/}(2005)Chakraborty, Balachandar \&
  Adrian]{ChakrabortyBalachandarAdrian2005}
{\sc \au{Chakraborty, P.}, \au{Balachandar, S.} \& \au{Adrian, R.~J.}}
  \yr{2005}  \at{On the relationships between local vortex identification
  schemes}.  \jt{Journal of Fluid Mechanics}  \bvol{535},  \pg{189--214}.

\bibitem[Chen {\em et~al.\/}(2006)Chen, Ecke, Eyink, Rivera, Wan \&
  Xiao]{ChenEckeEyinkRiveraWanXiao2006}
{\sc \au{Chen, S.}, \au{Ecke, R.~E.}, \au{Eyink, G.~L.}, \au{Rivera, M.},
  \au{Wan, M.} \& \au{Xiao, Z.}} \yr{2006}  \at{Physical mechanism of the
  two-dimensional inverse energy cascade}.  \jt{Physical Review Letters}
  \bvol{96},  \pg{084502}.

\bibitem[Chong {\em et~al.\/}(1990)Chong, Perry \&
  Cantwell]{ChongPerryCantwell1990}
{\sc \au{Chong, M.~S.}, \au{Perry, A.~E.} \& \au{Cantwell, B.~J.}} \yr{1990}
  \at{A general classification of three-dimensional flow fields}.  \jt{Physics
  of Fluids A: Fluid Dynamics}  \bvol{2}~(5),  \pg{765--777}.

\bibitem[Danilov \& Gurarie(2001)]{DanilovGurarie2001}
{\sc \au{Danilov, S.} \& \au{Gurarie, D.}} \yr{2001}  \at{Forced
  two-dimensional turbulence in spectral and physical space}.  \jt{Physical
  Review E}  \bvol{63}~(6),  \pg{061208}.

\bibitem[Eyink(2006{\natexlab{{\em a\/}}})]{Eyink2006}
{\sc \au{Eyink, G.~L.}} \yr{2006{\natexlab{{\em a\/}}}}  \at{Multi-scale
  gradient expansion of the turbulent stress tensor}.  \jt{Journal of Fluid
  Mechanics}  \bvol{549},  \pg{159--190}.

\bibitem[Eyink(2006{\natexlab{{\em b\/}}})]{Eyink2006b}
{\sc \au{Eyink, G.~L.}} \yr{2006{\natexlab{{\em b\/}}}}  \at{A turbulent
  constitutive law for the two-dimensional inverse energy cascade}.
  \jt{Journal of Fluid Mechanics}  \bvol{549},  \pg{191--214}.

\bibitem[Fang \& Ouellette(2016)]{FangOuellette2016}
{\sc \au{Fang, L.} \& \au{Ouellette, N.~T.}} \yr{2016}  \at{Advection and the
  efficiency of spectral energy transfer in two-dimensional turbulence}.
  \jt{Physical Review Letters}  \bvol{117}~(10),  \pg{104501}.

\bibitem[Farge \& Schneider(2015)]{FargeSchneider2015}
{\sc \au{Farge, M.} \& \au{Schneider, K.}} \yr{2015}  \at{Wavelet transforms
  and their applications to mhd and plasma turbulence: a review}.  \jt{Journal
  of Plasma Physics}  \bvol{81}.

\bibitem[Finn \& Apte(2013)]{FinnApte2013}
{\sc \au{Finn, J.} \& \au{Apte, S.~V.}} \yr{2013}  \at{Integrated computation
  of finite-time lyapunov exponent fields during direct numerical simulation of
  unsteady flows}.  \jt{Chaos}  \bvol{23},  \pg{013145}.

\bibitem[Frisch \& Sulem(1984)]{FrischSulem1984}
{\sc \au{Frisch, U} \& \au{Sulem, P.~L.}} \yr{1984}  \at{Numerical simulation
  of the inverse cascade in two-dimensional turbulence}.  \jt{The Physics of
  Fluids}  \bvol{27},  \pg{1921--1923}.

\bibitem[Hadjighasem {\em et~al.\/}(2017)Hadjighasem, Farazmand, Blazevski,
  Froyland \& Haller]{HadjighasemFarazmandBlazevskiFroylandHaller2017}
{\sc \au{Hadjighasem, A.}, \au{Farazmand, A.}, \au{Blazevski, D.},
  \au{Froyland, G.} \& \au{Haller, G.}} \yr{2017}  \at{A critical comparison of
  lagrangian methods for coherent structure detection}.  \jt{Chaos}  \bvol{27},
   \pg{053104}.

\bibitem[Haller(2005)]{Haller2005}
{\sc \au{Haller, G.}} \yr{2005}  \at{An objective definition of a vortex}.
  \jt{Journal of Fluid Mechanics}  \bvol{525},  \pg{1--26}.

\bibitem[Haller(2015)]{Haller2015}
{\sc \au{Haller, G.}} \yr{2015}  \at{Lagrangian coherent structures}.
  \jt{Annual Review of Fluid Mechanics}  \bvol{47},  \pg{137--162}.

\bibitem[Haller {\em et~al.\/}(2016)Haller, Hadjighasem, Farazmand \&
  Huhn]{HallerHadjighasemFarazmandHuhn2016}
{\sc \au{Haller, G.}, \au{Hadjighasem, A.}, \au{Farazmand, M.} \& \au{Huhn,
  F.}} \yr{2016}  \at{Defining coherent vortices objectively from the
  vorticity}.  \jt{Journal of Fluid Mechanics}  \bvol{795},  \pg{136--173}.

\bibitem[Haller \& Yuan(2000)]{HallerYuan2000}
{\sc \au{Haller, G.} \& \au{Yuan, G.}} \yr{2000}  \at{Lagrangian coherent
  structures and mixing in two-dimensional turbulence}.  \jt{Physica D:
  Nonlinear Phenomena}  \bvol{147},  \pg{352--370}.

\bibitem[Holmes {\em et~al.\/}(2012)Holmes, Lumley, Berkooz \&
  Rowley]{HolmesLumleyBerkoozRowley_TurbulenceCoherentStructuresDynamicalSystemsAndSymmetry_2012}
{\sc \au{Holmes, P.}, \au{Lumley, J.~L.}, \au{Berkooz, G.} \& \au{Rowley,
  C.~W.}} \yr{2012} {\em Turbulence, coherent structures, dynamical systems,
  and symmetry\/}, 2nd edn.  \publ{Cambridge University Press}.

\bibitem[Hua \& Klein(1998)]{HuaKlein1998}
{\sc \au{Hua} \& \au{Klein}} \yr{1998}  \at{An exact criterion for the stirring
  properties of nearly two-dimensional turbulence}.  \jt{Physica D}
  \bvol{113},  \pg{98--110}.

\bibitem[Hunt {\em et~al.\/}(1988)Hunt, Wray \& Moin]{HuntWrayMoin1988}
{\sc \au{Hunt, J. C.~R.}, \au{Wray, A.~A.} \& \au{Moin, P.}} \yr{1988}
  \at{Eddies, streams and convergence zones in turbulent flows}.
  \jt{Proceedings of the Summer Program 1988}  \pg{pp. 193--208}.

\bibitem[Jeong \& Hussain(1995)]{JeongHussain1995}
{\sc \au{Jeong, J.} \& \au{Hussain, F.}} \yr{1995}  \at{On the identification
  of a vortex}.  \jt{Journal of Fluid Mechanics}  \bvol{285},  \pg{69--94}.

\bibitem[Kelley {\em et~al.\/}(2013)Kelley, Allshouse \&
  Ouellette]{KelleyAllshouseOuellette2013}
{\sc \au{Kelley, D.~H.}, \au{Allshouse, M.~R.} \& \au{Ouellette, N.~T.}}
  \yr{2013}  \at{Lagrangian coherent structures separate dynamically distinct
  regions in fluid flows}.  \jt{Physical Review E}  \bvol{88},  \pg{013017}.

\bibitem[Kraichnan(1967)]{Kraichnan1967a}
{\sc \au{Kraichnan, R.~H.}} \yr{1967}  \at{Inertial ranges in two-dimensional
  turbulence}.  \jt{The Physics of Fluids}  \bvol{10},  \pg{1417--1423}.

\bibitem[Kraichnan(1971)]{Kraichnan1971}
{\sc \au{Kraichnan, R.~H.}} \yr{1971}  \at{Inertial-range transfer in two- and
  three-dimensional turbulence}.  \jt{Journal of Fluid Mechanics}  \bvol{47},
  \pg{525--535}.

\bibitem[Leith(1968)]{Leith1968}
{\sc \au{Leith, C.~E.}} \yr{1968}  \at{Diffusion approximation for
  two-dimensional turbulence}.  \jt{The Physics of Fluids}  \bvol{11},
  \pg{671--673}.

\bibitem[Leung(2011)]{Leung2011}
{\sc \au{Leung, S.}} \yr{2011}  \at{An eulerian approach for computing the
  finite time lyapunov exponent}.  \jt{Journal of Computational Physics}
  \bvol{230},  \pg{3500--3524}.

\bibitem[Liao \& Ouellette(2013)]{LiaoOuellette2013}
{\sc \au{Liao, Y.} \& \au{Ouellette, N.~T.}} \yr{2013}  \at{Spatial structure
  of spectral transport in two-dimensional flow}.  \jt{Journal of Fluid
  Mechanics}  \bvol{725},  \pg{281--298}.

\bibitem[Lilly(1971)]{Lilly1971}
{\sc \au{Lilly, D.~K.}} \yr{1971}  \at{Numerical simulation of developing and
  decaying two-dimensional turbulence}.  \jt{Journal of Fluid Mechanics}
  \bvol{45},  \pg{395--415}.

\bibitem[Maltrud \& Vallis(1993)]{MaltrudVallis1993}
{\sc \au{Maltrud, M.~E.} \& \au{Vallis, G.~K.}} \yr{1993}  \at{Energy and
  enstrophy transfer in numerical simulations of two-dimensional turbulence}.
  \jt{Physics of Fluids A: Fluid Dynamics}  \bvol{5}~(7),  \pg{1760--1775}.

\bibitem[Ohkitani(1991)]{Ohkitani1991}
{\sc \au{Ohkitani, K.}} \yr{1991}  \at{Wave number space dynamics of enstrophy
  cascade in a forced two-dimensional turbulence}.  \jt{Physics of Fluids A:
  Fluid Dynamics}  \bvol{3},  \pg{1598--1611}.

\bibitem[Okamoto {\em et~al.\/}(2007)Okamoto, Yoshimatsu, Schneider, Farge \&
  Kaneda]{OkamotoYoshimatsuSchneiderFargeKaneda2007}
{\sc \au{Okamoto, N.}, \au{Yoshimatsu, K.}, \au{Schneider, K.}, \au{Farge, M.}
  \& \au{Kaneda, Y.}} \yr{2007}  \at{Coherent vortices in high resolution
  direct numerical simulation of homogeneous isotropic turbulence: a wavelet
  viewpoint}.  \jt{Physics of Fluids}  \bvol{19},  \pg{115109}.

\bibitem[Okubo(1970)]{Okubo1970}
{\sc \au{Okubo, A.}} \yr{1970}  \at{Horizontal dispersion of floatable
  particles in the vicinity of velocity singularities such as convergences}.
  \jt{Deep Sea Research and Oceanographic Abstracts}  \bvol{17}~(3),
  \pg{445--454}.

\bibitem[Ouellette(2012)]{Ouellette2012}
{\sc \au{Ouellette, N.}} \yr{2012}  \at{On the dynamical role of coherent
  structures in turbulence}.  \jt{Comptes Rendus Physique}  \bvol{13},
  \pg{866--877}.

\bibitem[Paret \& Tabeling(1998)]{ParetTabeling1998}
{\sc \au{Paret, J} \& \au{Tabeling, P.}} \yr{1998}  \at{Intermittency in the
  two-dimensional inverse cascade of energy: experimental observations}.
  \jt{Physics of Fluids}  \bvol{10},  \pg{3126--3136}.

\bibitem[Pope(2000)]{Pope_TurbulentFlows_2000}
{\sc \au{Pope, S.~B.}} \yr{2000} {\em Turbulent flows\/}.  \publ{Cambridge
  University Press}.

\bibitem[Rowley {\em et~al.\/}(2009)Rowley, Mezi\'c, Bagheri, Schlatter \&
  Henningson]{RowleyMezicBagheriSchlatterHenningson2009}
{\sc \au{Rowley, C.~W.}, \au{Mezi\'c, I.}, \au{Bagheri, S.}, \au{Schlatter, P.}
  \& \au{Henningson, D.~S.}} \yr{2009}  \at{Spectral analysis of nonlinear
  flows}.  \jt{Journal of Fluid Mechanics}  \bvol{641},  \pg{115--127}.

\bibitem[Rutgers(1998)]{Rutgers1998}
{\sc \au{Rutgers, M.~A.}} \yr{1998}  \at{Forced 2d turbulence: experimental
  evidence of simultaneous inverse energy and forward enstrophy cascades}.
  \jt{Physical Review Letters}  \bvol{81},  \pg{2244--2247}.

\bibitem[Schmid(2010)]{Schmid2010}
{\sc \au{Schmid, P.~J.}} \yr{2010}  \at{Dynamic mode decomposition of numerical
  and experimental data}.  \jt{Journal of Fluid Mechanics}  \bvol{656},
  \pg{5--28}.

\bibitem[Scott(2007)]{Scott2007}
{\sc \au{Scott, R.~K.}} \yr{2007}  \at{Nonrobustness of the two-dimensional
  turbulent inverse cascade}.  \jt{Physical Review E}  \bvol{75},  \pg{046301}.

\bibitem[Staniforth \& C\^ot\'e(1991)]{StaniforthCote1991}
{\sc \au{Staniforth, A.} \& \au{C\^ot\'e, J.}} \yr{1991}  \at{Semi-lagrangian
  integration schemes and their application to environmental flows}.
  \jt{Monthly Review Weather}  \bvol{119}~(9),  \pg{2206--2223}.

\bibitem[Taira {\em et~al.\/}(2017)Taira, Brunton, Dawson, Rowley, Colonius,
  McKeon, Schmidt, Gordeyev, Theofilis \&
  Ukeiley]{TairaBruntonDawsonRowleyColoniusMcKeonSchmidtGordeyevTheofilisUkeiley2017}
{\sc \au{Taira, K.}, \au{Brunton, S.~L.}, \au{Dawson, S. T.~M.}, \au{Rowley,
  C.~W.}, \au{Colonius, T.}, \au{McKeon, B.~J.}, \au{Schmidt, O.~T.},
  \au{Gordeyev, S.}, \au{Theofilis, V.} \& \au{Ukeiley, L.~S.}} \yr{2017}
  \at{Modal analysis of fluid flows: an overview}.  \jt{AIAA Journal}
  \bvol{55},  \pg{4013--4041}.

\bibitem[Towne {\em et~al.\/}(2018)Towne, Schmidt \&
  Colonius]{TowneSchmidtColonius2018}
{\sc \au{Towne, A.}, \au{Schmidt, O.~T.} \& \au{Colonius, T.}} \yr{2018}
  \at{Spectral proper orthogonal decomposition and its relationship to dynamic
  mode decomposition and resolvent analysis}.  \jt{Journal of Fluid Mechanics}
  \bvol{847},  \pg{821--867}.

\bibitem[Vallgren(2011)]{Vallgren2011}
{\sc \au{Vallgren, A.}} \yr{2011}  \at{Infrared reynolds number dependency of
  the two-dimensional inverse energy cascade}.  \jt{Journal of Fluid Mechanics}
   \bvol{667},  \pg{463--473}.

\bibitem[Weiss(1991)]{Weiss1991}
{\sc \au{Weiss, J.}} \yr{1991}  \at{The dynamics of enstrophy transfer in
  two-dimensional hydrodynamics}.  \jt{Physica D}  \bvol{48},  \pg{273--294}.

\bibitem[Xiao {\em et~al.\/}(2009)Xiao, Wan, Chen \&
  Eyink]{XiaoWanChenEyink2009}
{\sc \au{Xiao, Z.}, \au{Wan, M.}, \au{Chen, S.} \& \au{Eyink, G.~L.}} \yr{2009}
   \at{Physical mechanism of the inverse energy cascade of two-dimensional
  turbulence: a numerical investigation}.  \jt{Journal of Fluid Mechanics}
  \bvol{619},  \pg{1--44}.

\bibitem[Yadav {\em et~al.\/}(2021)Yadav, Cameron \&
  Solanki]{YadavCameronSolanki2021}
{\sc \au{Yadav, N.}, \au{Cameron, R.~H.} \& \au{Solanki, S.~K.}} \yr{2021}
  \at{Vortex flow properties in simulations of solar plage region: evidence for
  their role in chromospheric heating}.  \jt{Astronomy and Astrophysics}
  \bvol{645},  \pg{A3}.

\bibitem[Yoshimatsu {\em et~al.\/}(2009)Yoshimatsu, Kondo, Schneider, Okamoto,
  Hagiwara \& Farge]{YoshimatsuKondoSchneiderOkamotoHagiwaraFarge2009}
{\sc \au{Yoshimatsu, K.}, \au{Kondo, Y.}, \au{Schneider, K.}, \au{Okamoto, N.},
  \au{Hagiwara, H.} \& \au{Farge, M.}} \yr{2009}  \at{Wavelet-based coherent
  vorticity sheet and current sheet extraction from three-dimensional
  homogeneous magnetohydrodynamic turbulence}.  \jt{Physics of Plasmas}
  \bvol{16},  \pg{082306}.

\bibitem[Zhou {\em et~al.\/}(1999)Zhou, Adrian, Balachandar \&
  Kendall]{ZhouAdrianBalachandarKendall1999}
{\sc \au{Zhou, J.}, \au{Adrian, R.~J.}, \au{Balachandar, S.} \& \au{Kendall,
  T.~M.}} \yr{1999}  \at{Mechanisms for generating coherent packets of hairpin
  vortices in channel flow}.  \jt{Journal of Fluid Mechanics}  \bvol{387},
  \pg{353--396}.

\end{thebibliography}
\end{document}